\documentclass[12pt]{iopart}

\usepackage{graphicx}
\usepackage{color}
\usepackage{bm}
\usepackage{amssymb}
 \usepackage{cite}

\begin{document}

\title[Thermodynamic and Logical Reversibilities Revisited]{Thermodynamic and Logical Reversibilities Revisited}

\author{Takahiro Sagawa}

\address{Department of Basic Science, The University of Tokyo, 3-8-1 Komaba, Meguro-ku, Tokyo 153-8902, Japan}
\ead{sagawa@noneq.c.u-tokyo.ac.jp}
\begin{abstract}
We review and investigate the general theory of thermodynamics of computation, and derive the fundamental inequalities that set the lower bounds of the work requirement and the heat emission during a computation.  These inequalities constitute the generalized Landauer principle, where the information contents are involved in the second law of thermodynamics.  We discuss in detail the relationship between the thermodynamic and logical reversibilities; the former is related to the entropy production in the total system including a heat bath, while the latter is related to the entropy change only in the logical states of the memory.  In particular, we clarify that any logically irreversible computation can be performed in a thermodynamically reversible manner in the quasi-static limit, which does not contradict the conventional Landauer principle. Our arguments would serve as the theoretical foundation of thermodynamics of computation in terms of modern statistical physics.
\end{abstract}

\maketitle

\section{Introduction}

The investigation of thermodynamics of computation and information processing  predates the era of computer technology~\cite{Demon}.  
From the viewpoint of fundamental physics, this research area is closely related to the foundation of the second law of thermodynamics, which dates back to Maxwell's thought experiment on ``Maxwell's demon'' in the ninteenth century~\cite{Maxwell}.   Several decades later, Szilard made the first crucial step toward the quantitative understanding of the relationship between information and thermodynamics~\cite{Szilard}.  In his thought experiment with a  single-particle heat engine, which is called the Szilard engine, the demon can extract $k_{\rm B}T \ln 2$ of work from a single heat bath through feedback control by using one bit of information, where $k_{\rm B}$ is the Boltzmann constant and $T$ is the temperature of the bath.
By developing Szilard's observation, Brillouin investigated the relationship between thermodynamic and informational entropies~\cite{Brillouin1,Brillouin2}.
Recently, a kind of the Szilard engine have been experimentally demonstrated for the first time~\cite{Toyabe} with a small thermodynamic engine that works at the level of thermal fluctuations.

Another prominent observation was made by Landauer~\cite{Landauer,Landauer2}.  He pointed out that, to erase one bit of information from a memory, at least $k_{\rm B}T \ln 2$ of heat should be emitted into a heat bath and the same amount of work is needed.  This has been referred to as the Landauer principle.  
Bennett also discussed thermodynamics of computation  by considering  the concept of logical reversibility and its relationship to thermodynamics~\cite{Bennett0,Bennett1,Bennett2}.
Moreover, Zurek discussed thermodynamics of computation in terms of algorithmic complexity~\cite{Zurek1,Zurek2}.
Later, the Landauer principle has been studied in various aspects~\cite{Plenio,Maruyama}; it has been derived in terms of statistical mechanics in some setups~\cite{Shizume,Piechocinska,Esposito2}, has been demonstrated in concrete systems theoretically~\cite{Lutz1,Lutz2,Lambson,Aurell,Diana1} and experimentally~\cite{Berut,Berut2}, has sparked up intense discussions on its validity~\cite{Goto,Allahverdyan3,Horhammer,Vaccaro,Maroney0,Norton,Barkeshli,Maroney,Turgut,Sagawa-Ueda2}.


In a rather different context, the recent advancements in nonequilibrium statistical mechanics have revealed a fundamental aspect of the second law of thermodynamics~\cite{Jarzynski1,Crooks1,Crooks2,Jarzynski2,Seifert,Kawai,Jarzynski3,Crooks3,Kurchan,Tasaki0,Campisi,Sagawa3}.
This has enabled us to more quantitatively and comprehensively understand thermodynamics of computation and information processing. 
In terms of theory, the second law of thermodynamics and the fundamental nonequilibrium relations (e.g, the fluctuation theorem and the Jarzynski equality) have been generalized to information processing processes such as measurement, information erasure, and feedback control~\cite{Nielsen,Touchette,Touchette2,Sagawa-Ueda1,Cao2,Jacobs,Sagawa-Ueda3,Suzuki,Horowitz1,Horowitz2,Morikuni,Abreu,Horowitz3,Sagawa-Ueda4,Abreu2,Still,Vedral,Sagawa-Ueda2012,Granger,Ito2,Deffner,Tasaki,HSP,Sagawa-Ueda-NJP}, where information contents and thermodynamic quantities are treated on an equal footing. 
In particular, the concept of thermodynamic reversibility in the presence of feedback control has been established~\cite{Horowitz2,Sagawa-Ueda4}, and several concrete models of thermodynamically reversible information processing have been proposed~\cite{Jacobs,Horowitz2,Abreu,Horowitz3,Sagawa-Ueda4,HSP}.
More recently, a variety of autonomous Maxwell's demons has attracted much attention~\cite{
Mandal,Barato,Strasberg,Mandal2,Barato2}.
In terms of experiment, a generalized Jarzynski equality with feedback control has been verified~\cite{Toyabe}.  
The Jarzynski equality for information erasure has also been investigated experimentally~\cite{Berut2}.
In light of these advancements, we are now in the position to revisit and to further clarify the fundamental concepts in thermodynamics of computation and information processing.

In this paper, we review and investigate the long-standing fundamental problems in thermodynamics of computation, shedding new light on the celebrated Landauer principle from the viewpoint of  modern statistical physics.
In particular, we clarify the fundamental relationship between the thermodynamic and logical reversibilities.
Moreover, on the basis of the second law of thermodynamics, we derive universal thermodynamic inequalities that set the lower bounds of the work requirement and the heat emission during a computation.
Our observation does not contradict the conventional Landauer principle, but extends it to much broader class of  memories that perform computation. 
While we assume  that a memory  is a classical system, the generalization of our arguments to the quantum cases is straightforward~\cite{Sagawa3}.

This paper is organized as follows.  In Sec.~2, we discuss the concept of thermodynamic reversibility in terms of the total entropy production in the whole system including a heat bath.  In Sec.~3, we discuss the concept of logical reversibility, and illustrate several typical examples of reversible and irreversible computations.   In Sec.~4, we discuss the conventional Landauer principle, and clarify the  relationship between thermodynamic and logical reversibilities with the standard setup of the Landauer principle.  In Sec.~5, we formulate the general setup of thermodynamic computation, and derive the second law of thermodynamics for computational processes.
In Sec.~6, we discuss the situation that there are two memories; this setup enables us to analyze measurement and feedback, which constitute the typical situation of Maxwell's demon.  In Sec.~7, we conclude this paper.

\section{Thermodynamic Reversibility}

In this section, we clarify the concept of thermodynamic reversibility.
We consider a time evolution of a thermodynamic system in the presence of heat bath(s).
We assume that the system may be driven from equilibrium by changing external parameters such as the volume of the gas.
First of all, we roughly characterize the thermodynamic reversibility according to the standard definition in thermodynamics~\cite{Callen,Prigogine}:\\
\\
{\it A physical process is thermodynamically reversible if and only if its time-reversal is not prohibited by the second law of thermodynamics.  Otherwise, the process is thermodynamically irreversible.}\\
\\
In more precise, we need to consider the ensemble of the system, because, from the microscopic point of view,  the dynamics of the phase-space point of the system is stochastic due to thermal fluctuations. The thermodynamic property of the system is described by the probability distribution on the phase space.  The thermodynamic reversibility can then be characterized as follows~\cite{Jarzynski1,Crooks1,Crooks2,Jarzynski2,Seifert,Kawai,Crooks3}:\\
\\
{\it A physical process is thermodynamically reversible if and only if the time evolution of the probability distribution in the process can be time-reversed, where the change of the external parameters is also time-reversed, and the signs of the amounts of the work and the heat are changed.}\\
\\
A process can become thermodynamically reversible  in the quasi-static limit, where the change of the external parameters is much slower than the relaxation time of the system, and the state of the system can always be regarded in thermal equilibrium during the process.
For example, the quasi-static and isothermal compression and expansion of the gas in a box are both thermodynamically reversible, as they are the time-reversal with each other.
We note that all quasi-static processes are not necessarily reversible; there may be quasi-static but  irreversible  processes such as a weak constant of two baths with different temperatures~\cite{Callen}.

To characterize the thermodynamic reversibility in a more quantitative way, we discuss the concept of  entropy production.
For simplicity, we assume that there is a single heat bath at inverse temperature $\beta := (k_{\rm B}T)^{-1}$.  The generalization of the following arguments to the cases with multiple baths is straightforward.
Let $\mathcal Y$ be the phase space of the system, $y, y' \in \mathcal Y$ be the initial and final phase-space points, and $P[y]$ and $P'[y']$ be their probabilities. 
We consider the Shannon entropy of the probability distribution on the phase space~\cite{Shannon,Cover-Thomas}.  The initial and final entropies are respectively given by
\begin{equation}
S := - \int_{y \in \mathcal Y} dy P[y] \ln P[y], \ \ S' := - \int_{y' \in \mathcal Y} dy' P'[y'] \ln P'[y'].
\end{equation}
We denote as $Q$ the  average heat absorbed by the system from the heat bath during the dynamics.
The total entropy production  in the relevant total system (i.e., the whole ``universe'') including the bath is then given by~\cite{Crooks1,Crooks2,Jarzynski2,Seifert}
\begin{equation}
\Delta S_{\rm tot} := \Delta S -\beta Q,
\label{total_entropy_production}
\end{equation}
where  $\Delta S := S' - S$ is the change in the Shannon entropy of the system.  We note that $-\beta Q$ is  regarded as the change in the entropy of the heat bath.  

The second law of thermodynamics can then be expressed as~\cite{Crooks1,Crooks2,Jarzynski2,Seifert,Callen,Prigogine}
\begin{equation}
\Delta S_{\rm tot} \geq 0,
\label{second1}
\end{equation}
or equivalently,
\begin{equation}
\Delta S \geq \beta Q.
\label{second1_1}
\end{equation}
We stress that inequality~(\ref{second1})  holds for any initial and final nonequilibrium distributions of the system (i.e., for any $P[y]$ and $P'[y']$).
Inequality~(\ref{second1}) can be derived on the basis of nonequilibrium statistical mechanics~\cite{Sagawa3}.  In fact, inequality (\ref{second1})  is a straightforward consequence of the fluctuation theorem~\cite{Crooks1,Crooks2,Jarzynski2,Seifert}.

On the basis of inequality~(\ref{second1}), the thermodynamic reversibility can be characterized in terms of the entropy production~\cite{Callen,Prigogine}:  \\
\\
{\it  A physical process is thermodynamically reversible if  the equality in (\ref{second1}) is achieved (i.e., the total entropy production is zero). }\\
\\
In fact, if $\Delta S_{\rm tot} > 0$ holds in a process, its time-reversal is impossible because of $\Delta S_{\rm tot} < 0$.  
The thermodynamically reversible condition $\Delta S_{\rm tot} = 0$ can be achieved in the quasi-static limit, where the system is always in thermal equilibrium during the process.

We stress that a process can be thermodynamically reversible even if there is an entropy transfer from the system to the bath or vice versa.  If the amount of the increase/decrease in entropy in the system is the same as the amount of the decrease/increase in the bath, the total amount of the entropy increase in the whole universe is  zero, where the equality in (\ref{second1}) is achieved  and the process is thermodynamically reversible.  

As a simple example, we consider a quasi-static isothermal expansion of an ideal gas with $N$ particles.  Starting from a thermal equilibrium state, we expand the gas isothermally and quasi-statically, doubling its volume. The entropy of the system is then increased by $N \ln 2$, and the heat emission to the bath is given by $-  Q = \beta^{-1} N \ln 2$ (i.e.,  the entropy of the bath is decreased by $N \ln 2$).  Therefore, the total entropy production is given by
\begin{equation}
\Delta S_{\rm tot} = ( -N \ln 2) +  N \ln 2 = 0,
\end{equation}
which implies that  the quasi-static isothermal expansion is thermodynamically reversible.  Similarly, the quasi-static isothermal compression is also thermodynamically reversible.

We now discuss the relationship between the foregoing thermodynamic perspective and the microscopic perspective based on the reversible Hamiltonian dynamics.
One of the crucial progresses in modern statistical physics is that it has succeeded to reconcile these two perspectives~\cite{Jarzynski1,Jarzynski2,Kawai}.
In fact, the second law (\ref{second1}) can be derived on the basis of the fluctuation theorem, where one needs essentially  only two assumptions: the microscopic dynamics satisfies the Liouville theorem, and the initial distribution of the bath is the canonical distribution~\cite{Jarzynski2}.  
A crucial observation here is that the Shannon entropy on the whole phase space of the system and the bath is different from the thermodynamic entropy in the whole universe; the former is conserved in any Hamiltonian dynamics, while the latter is increased in thermodynamically irreversible processes even if the underlying microscopic Hamiltonian dynamics is reversible.

The change in the thermodynamic entropy is equal to the relative entropy (i.e., the Kullback-Leibler divergence) between the probability distribution in the final state and a reference probability distribution such as the canonical distribution~\cite{Jarzynski3}.
We again stress that the thermodynamic entropy is not equivalent to the Shannon entropy of the total system including the system and the bath.
Correspondingly, we have not considered the Shannon entropy in the bath in the foregoing argument, but regarded the heat transfer $-\beta Q$ as the entropy change in the bath, which is consistent with the picture that the relative entropy corresponds to the entropy production in the whole universe~\cite{Jarzynski2,Jarzynski3,Sagawa3}. 
The thermodynamic irreversibility (i.e., the positive entropy production in the whole universe) can also be characterized by the gap between the probability distributions of microscopic trajectories in the forward process and the backward one~\cite{Kawai,Crooks3,Jarzynski3}.
We note that the same argument is also valid for quantum systems that obey unitary dynamics~\cite{Kurchan,Tasaki0,Campisi,Sagawa3}, by replacing the Shannon entropy by the von Neumann one.

The concept of thermodynamic reversibility discussed in this manuscript is consistent with the above-mentioned observation based on the fluctuation theorem, and therefore has a rigid theoretical foundation that is consistent with the microscopic reversible physics.
In terms of the fluctuation theorem, $\Delta S_{\rm tot} = 0$ holds if and only if the probability distribution of the trajectories in the phase space is the same as that of the time-reversed trajectories, which has been explicitly discussed in, for example, Ref.~\cite{Crooks3,Jarzynski3}.

We now assume that the initial and final distributions are the canonical distributions:
\begin{equation}
P[y] = e^{\beta (F - E[y])}, \ P'[y'] = e^{\beta (F'- E'[y'])},
\end{equation}
where $E[y]$ ($E'[y']$)  and $F$ ($F'$) are the initial (final) energy  and the initial (final) free energy of the system, respectively.  In this case, we obtain
\begin{equation}
S  = \beta (E   - F  ), \ \ S' = \beta (E'  - F' ),
\end{equation}
where  $E$ ($E'$) is the ensemble average of the initial (final)  energy. Therefore,
\begin{equation}
\Delta S = \beta (\Delta E  - \Delta F), 
\end{equation}
where $\Delta E := E' -E$ and $\Delta F := F' -F$.
Let $W$ be the average work performed on the system during the process.  The total entropy production is then given by
\begin{equation}
\Delta S_{\rm tot} = \beta (W  - \Delta F),
\end{equation}
where we used the first law of thermodynamics:
\begin{equation}
W  =  \Delta E  -  Q.
\end{equation}
Therefore, the second law~(\ref{second1}) reduces to
\begin{equation}
W \geq \Delta F,
\label{second2}
\end{equation}
where the equality can be achieved in the quasi-static limit.

If the final distribution is different from the canonical distribution, we can show that
\begin{equation}
\Delta S_{\rm tot} \leq \beta ( W - \Delta F),
\label{S_F_noneq}
\end{equation}
where the final free energy is given by the equilibrium one corresponding to the final Hamiltonian.
In this case, we again obtain inequality~(\ref{second2}).
We note that the equality in (\ref{S_F_noneq}) is achieved if the final distribution is given by the canonical distribution.

\section{Logical Reversibility}

In this section, we discuss the concept of logical reversibility, and introduce the logical entropy that is decreased by logically irreversible computation.

\subsection{Definition and examples}

We consider a memory that has several logical states, in which information is stored.
A computational process is performed by changing an input logical state into output one with a certain algorithm. 
Let $\mathcal M$ and $\mathcal M'$ be the sets of input and output logical states, respectively.
We assume that these are finite sets.
For example, if the input consists of $n$ bits,  we set $\mathcal M = \{ 0,1 \}^n$.
We note that the logical states do not have one-to-one correspondence to the physical phase space, as discussed in Sec.~4 in detail.

We next formulate (deterministic) computational processes.
In this paper,  a computational process $\hat C$ is defined as a map  
\begin{equation}
\hat C: \mathcal M \to \mathcal M'.
\end{equation}
We also call $\hat C$ as a gate.
We note that $\hat C$ is not necessarily a surjection.

In order to discuss the logical reversibility and its relationship to thermodynamics,  we do not need the detailed characterization  of  computable functions  in terms of computability theory~\cite{Moore}; the following argument is  applicable to any map $\hat C : \mathcal M \to \mathcal M'$.

We now define the logical reversibility~\cite{Landauer2,Bennett0,Bennett1,Bennett2}: \\
\\
{\it A computational process $\hat C$ is logically reversible if and only if it is an injection.
In other words, $\hat C$ is logically reversible if and only if, for any output logical state, there is a unique input logical state.
Otherwise,  $\hat C$ is logically irreversible.}\\
\\
We note that, if $\mathcal M' = \mathcal M$, a computational process is reversible if and only if it is a bijection (i.e., an injection and a surjection).
In general, a computational process is reversible if and only if we can precisely estimate the input state from the output state.
In fact, if a computational process is reversible and $\hat C$ is an injection, we can define the reversed computational process
\begin{equation}
\hat C^{-1}: \mathcal M' \to \mathcal M,
\end{equation}
where the domain of $\hat C^{-1}$ is given by the image of $\hat C$ denoted as $\hat C (\mathcal M) \subset \mathcal M'$.

We next discuss several simple examples of computation.
We first consider the case that both of the input and output are one bit so that $\mathcal M = \mathcal M' = \{ 0,1 \}$.
A simple example of reversible gate is NOT, which is defined as
\begin{equation}
0 \mapsto 1, \ \ 1 \mapsto 0.
\end{equation}
This is clearly a bijection, and the reversal of NOT is also NOT.
A simple example of irreversible gate is the information erasure that is referred to as ERASE:
\begin{equation}
0 \mapsto 0, \ \ 1 \mapsto 0.
\end{equation}
This is not a bijection, as the logical state is always $0$ after the computation; we cannot estimate the input state from the output state.

We next consider  the case that both of the input and output are two bits so that $\mathcal M = \mathcal M' = \{ 0,1 \}^2 = \{ 00, 01, 10, 11 \}$.
A simple example of reversible gate is CNOT (controlled-NOT), which is defined as
\begin{equation}
00 \mapsto 00, \ \ 01 \mapsto 01, \ \ 10 \mapsto 11, \ \ 11 \mapsto 10,   
\label{CNOT}
\end{equation}
where the first (left) bit is the control bit and the second (right) bit is the target bit.  If the control gate is $1$, CNOT behaves as NOT on the target bit.  Otherwise, CNOT is just identity.  CNOT is a bijection and its reversal is also CNOT.

CNOT can be used for a measurement (i.e., the copy of information); if the input of the target bit is $0$,  the input of the control bit is copied to the output of the target bit by CNOT:
\begin{equation}
00 \mapsto 00,  \ \ \ 10 \mapsto 11.
\label{CNOT_measurement}
\end{equation}
CNOT can also be used for feedback control, where the control bit is the feedback controller and  the target bit is to be controlled. 
In the case of feedback control, we exchange the roles of the bits from the case of measurement~(\ref{CNOT_measurement}); we regard the first (left) bit as the target bit and the second (right) bit as the control bit for feedback control.
Before the feedback, the input states of the two bits are assumed to be the same, which is realized after the measurement. The target bit is then flipped if the control bit is $1$: 
\begin{equation}
00 \mapsto 00, \ \ 11 \mapsto 01,
\label{CNOT_feedback}
\end{equation}
where the output of the target bit is $0$ irrespective of its input.


We next consider the case that the input is two bit and the output is one bit, where $\mathcal M =  \{ 0,1 \}^2$ and $\mathcal M' = \{ 0, 1 \}$.  In this case, any computational process is irreversible.  In fact, the numbers of  the elements in $\mathcal M$ and $\mathcal M'$  are $4$ and $2$, respectively, and therefore any map from $\mathcal M$ to $\mathcal M'$ cannot be an injection.  An example of such an irreversible gate is XOR, which is defined as
\begin{equation}
00 \mapsto 0, \ \ 01 \mapsto 1, \ \ 10 \mapsto 1, \ \ 11 \mapsto 0.
\end{equation}

\subsection{Reversible extension}

We next show that any irreversible computation can be extended to a reversible computation, which has been discussed by Bennett in detail~\cite{Bennett0}.  In fact, we can show the following theorem: \\
\\
{\it For any $\hat C : \mathcal M \to \mathcal M'$, there exist a finite set $\mathcal M''$ and a map $\hat C_{\rm ex}: \mathcal M \to \mathcal M' \times \mathcal M''$,  such that $\hat C_{\rm ex}$ is logically reversible and the restriction of $\hat C_{\rm ex}$ on $\mathcal M \to \mathcal M'$ is equivalent to $\hat C$.}  \\
\\
Here, $\mathcal M''$ can be regarded as an ancilla or an environment. 
We call $\hat C_{\rm ex}$ the reversible extension of $\hat C$.
We note that the reversible extension is not unique.

Before the proof of the theorem, we illustrate the reversible extension of ERASE.  Let $\mathcal M'' = \{ 0,1 \}$.  We then define $\hat C_{\rm ex} : \{ 0,1 \} \to \{ 0,1 \}^2$ by
\begin{equation}
0 \mapsto 00, \ \ \   1 \mapsto  01,
\label{ERASE_ex}
\end{equation}
where the first (left) bit of the output is in $\mathcal M$.  $\hat C_{\rm ex}$ is clearly an injection and therefore reversible.  Moreover, its restriction on $\mathcal M \to \mathcal M'$ is equivalent to ERASE.  Intuitively, this extension describes that the erased information can be kept in ancilla $\mathcal M''$ (i.e., can remain in the environment).

We now show a simple constructive proof of the theorem: we set $\mathcal M'' = \mathcal M$ and define $\hat C_{\rm ex}$ as
\begin{equation}
\hat C_{\rm ex} (m) = (\hat C(m), m ) \in \mathcal M' \times \mathcal M'',
\label{reversible_ex}
\end{equation}
which satisfies the condition of the theorem.  (Q.E.D.)    

The extension of ERASE~(\ref{ERASE_ex}) is a special case of  extension~(\ref{reversible_ex}).  We note that  extension~(\ref{reversible_ex}) may be redundant in general.   For example, in the case of XOR, the extension~(\ref{reversible_ex}) is given by
\begin{equation}
00 \mapsto 000, \ \ 01 \mapsto 101, \ \ 10 \mapsto 110, \ \ 11 \mapsto 011,
\end{equation}
where $\mathcal M'' = \{ 0, 1 \}^2$.  However, there is a simpler reversible extension of XOR:
\begin{equation}
00 \mapsto 00, \ \ 01 \mapsto 10, \ \ 10 \mapsto 11, \ \ 11 \mapsto 01,
\end{equation}
where $\mathcal M'' = \{ 0 , 1 \}$.

\subsection{Entropy change in computation}

We next consider the concept of logical entropy, which is defined by the Shannon entropy of the logical states.

We consider probability distribution on $\mathcal M$ (i.e., $P[m]$ for $m \in \mathcal M$) with  $\sum_{m \in \mathcal M} P[m] = 1$.  After computation $\hat C$, the probability distribution on $\mathcal M'$ (i.e., $P'[m']$ for $m' \in \mathcal M'$) is given by
\begin{equation}
P' [m'] = \sum_{m: \ \hat C (m) = m'} P[m],
\label{prob_com1}
\end{equation}
where the sum in the right-hand side is taken over  $m$ satisfying $\hat C (m ) = m'$.
If the computation is reversible, Eq.~(\ref{prob_com1}) reduces to
\begin{equation}
P' [m'] =  P[\hat C^{-1} (m') ] \ \  ({\rm if} \ m' \in \hat C (\mathcal M)), \ \ \ P' [m'] = 0 \ \ {\rm (otherwise)},
\label{prob_com2}
\end{equation}
which describes the conservation of probability.

We now define the initial and final logical entropies by~\cite{Shannon,Cover-Thomas} 
\begin{equation}
H (\mathcal M) := - \sum_{m \in \mathcal M} P[m] \ln P[m],
\end{equation}
\begin{equation}
H' (\mathcal M') := - \sum_{m' \in \mathcal M'} P'[m'] \ln P'[m'].
\end{equation}
If the computation is logically reversible, the logical entropy does not change.  In fact, by using Eq.~(\ref{prob_com2}), we obtain
\begin{equation}
\eqalign{
H' (\mathcal M') &= - \sum_{m' \in \hat C (\mathcal M)} P[\hat C^{-1} (m') ] \ln P[\hat C^{-1} (m') ] \\
&= - \sum_{m \in \mathcal M} P[m] \ln P[m] = H (\mathcal M).
}
\end{equation}
In contrast, if the computation is logically irreversible, the logical entropy states is decreased:
\begin{equation}
H(\mathcal M) \geq H'(\mathcal M').
\label{Shannon_decrease}
\end{equation}
In fact, by noting that $P'[m'] = \sum_{m : \ \hat C (m) = m'} P[m]$ (i.e., $P[m] / P'[m']$ is a probability distribution over $m$'s that satisfy $\hat C (m) = m'$), we obtain
\begin{equation}
H(\mathcal M) - H'(\mathcal M')  = - \sum_{m'} P' [m'] \sum_{m : \  \hat C (m ) = m'} \frac{P[m]}{P'[m']} \ln \frac{P[m]}{P'[m']} \geq 0.
\end{equation}
We note that inequality~(\ref{Shannon_decrease}) is a special case of the data processing inequality in information theory~\cite{Cover-Thomas}.

In the case of the reversible extension~(\ref{reversible_ex}), the logical entropy of the extended logical states does not change: $H(\mathcal M ) = H' (\mathcal M' \times \mathcal M'')$.  From the subadditivity of the Shannon entropy, we have
\begin{equation}
H'(\mathcal M' \times \mathcal M'') \leq H'(\mathcal M') + H'( \mathcal M''),
\label{subadditivity}
\end{equation}
and therefore, together with inequality (\ref{Shannon_decrease}), 
\begin{equation}
H'(\mathcal M')  \leq H(\mathcal M ) \leq H'(\mathcal M') + H'( \mathcal M'').
\end{equation}

We discuss simple examples with $\mathcal M = \mathcal M' = \{ 0,1 \}$.  Let $p := P[0]$ and $p' := P' [0]$ be the probabilities of the input $0$ and output $0$, respectively.  In the case of NOT, $p' = 1-p$ and therefore $H(\mathcal M) = - p \ln p - (1-p) \ln (1-p) = H'(\mathcal M')$.  In the case of ERASE, $p' = 1$ for any $p$.  Therefore, $H(\mathcal M) =  - p \ln p - (1-p) \ln (1-p)$ and $H' (\mathcal M') = 0$, which implies that the logical entropy is decreased by $H(\mathcal M)$ by information erasure. 
In the case of the reversible extension of erasure~(\ref{ERASE_ex}),  the equality in~(\ref{subadditivity}) is achieved with $ H'(\mathcal M')  = 0$ and $H'( \mathcal M'') = - p \ln p - (1-p) \ln (1-p)$.  Therefore, the decrease in the entropy of the memory  is compensated for by the increase in the entropy of the ancilla or environment, such that the total entropy is conserved.

\section{Conventional Landauer Principle}

In this section, we discuss the conventional Landauer principle~\cite{Landauer}, and clarify the relationship between the thermodynamic and logical reversibilities in the standard setup of information erasure.

We consider the information erasure of one bit of information from a memory in the presence of a single heat bath at inverse temperature $\beta$.
As a simple and conventional setup, we consider a binary symmetric potential as a physical model of the memory (Fig.~1 (a))~\cite{Landauer,Landauer2,Bennett1}, where the height of the barrier is assumed to be much larger than the thermal fluctuation.
If the particle is in the left (right) well, the logical state is  ``$0$'' (``$1$'').  
An idealized  model of the double-well memory is shown in Fig.~1 (b), where the left (right) box corresponds to the left (right) well, and the wall at the center of the box corresponds to the barrier of the binary potential.
We note that these two models are  not completely equivalent.  
In fact, the width of the barrier is finite in the double-well memory but infinitely small in the two-box memory.
We also note that the height of the barrier in the double-well memory can be very large but is still finite theoretically, while the barrier in the two-box memory can be regarded as perfectly impenetrable.

\begin{figure}[htbp]
\begin{center}
\includegraphics[width=75mm]{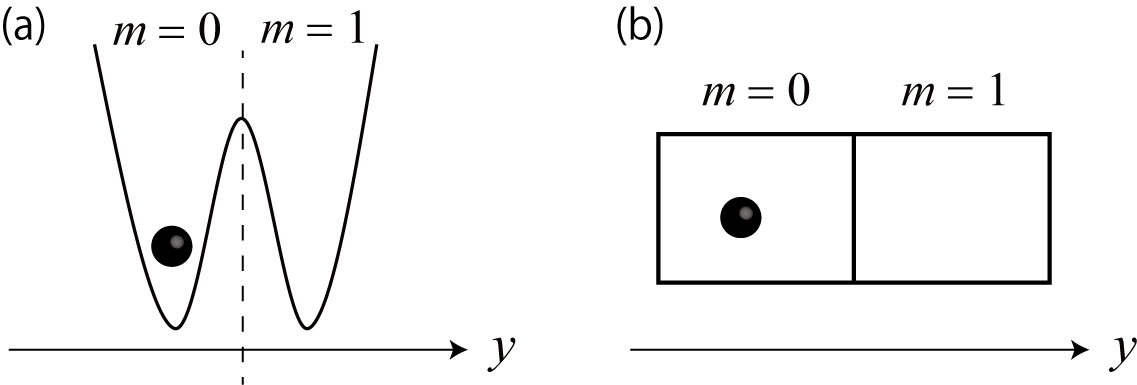}
\end{center}
\caption{(a) Schematic of a memory with a binary symmetric potential.  If the particle is in the left (right) well, the logical state is $0$ ($1$).  (b) A model of the memory with two boxes, which is an idealization of the binary-potential memory (a).  If the particle is in the left (right) box, the logical state is $0$ ($1$).}
\end{figure}

Before the erasure, the probability of ``$0$''  and ``$1$'' are assumed to be  equally $1/2$.  After the  erasure, the particle is in ``$0$'' with unit probability.  
The logical entropy is $\ln 2$ before the erasure, while it is $0$ after the erasure.  Therefore, the logical entropy changes by $\Delta H = -\ln 2$ during the erasure.  
We assume that the initial probability distribution in the memory is thermal equilibrium, and that the final probability distribution is in conditional thermal equilibrium only in the well of ``$0$.''

The crucial observation in the Landauer principle~\cite{Landauer,Landauer2,Bennett0,Bennett1,Bennett2} is that the logical entropy must be included as a part of the total entropy, and be treated on an equal footing with the thermodynamic entropy.
The change in the total entropy of the memory is then given by
\begin{equation}
\Delta S = \Delta H = - \ln 2,
\end{equation}
and therefore, the second law~(\ref{second1}) reduces to
\begin{equation}
-  Q \geq \beta^{-1} \ln 2,
\label{Landauer1}
\end{equation}
where $-  Q$ is the heat that is emitted to the heat bath during the erasure. 
Inequality~(\ref{Landauer1}) is the conventional Landauer principle and  the right-hand side is called the Landauer bound.  Inequality~(\ref{Landauer1}) implies that the decrease in the entropy by $\ln 2$ in the memory must be compensated for by the increase in the entropy of the bath by at least $\ln 2$, which is accompanied by the inevitable heat emission of $- Q = \beta^{-1} \ln 2$.
We note that the increase in the entropy of the bath corresponds to the  increase in the entropy of ancilla $\mathcal M''$ in terms of the reversible erasure~(\ref{ERASE_ex}) in Sec.~3.2.
The Landauer principle can then be summarized as follows: \\
\\
{\it A positive amount of the heat emission is inevitable during the logically irreversible information erasure.}\\
\

We next consider the work needed for the erasure.
Since  the internal energy of the memory does not change during the erasure, we have $- Q = W$ from the first law of thermodynamics.  Therefore, we obtain the minimal work needed for the information erasure:
\begin{equation}
W \geq \beta^{-1} \ln 2.
\label{Landauer2}
\end{equation}
Inequality~ (\ref{Landauer2}) is also referred to as the Landauer principle.

If the information erasure is quasi-static, the equalities in (\ref{Landauer1}) and (\ref{Landauer2}) can be achieved as
\begin{equation}
 -  Q  = \beta^{-1}  \ln 2,  \ \  W = \beta^{-1} \ln 2.
\end{equation}
A concrete protocol of such a quasi-static erasure will be discussed below.
Therefore, the total entropy production is given by
\begin{equation}
\Delta S_{\rm tot} = (-\ln 2) - (- \ln 2) = 0,
\end{equation}
which implies that the quasi-static erasure is thermodynamically reversible, while it is logically irreversible.
We now conclude that:\\
\\
{\it The logically irreversible erasure can be performed  in a thermodynamically reversible manner in the quasi-static limit.}\\
\\
This does not contradict the conventional Landauer principle. 
In fact, the logical reversibility is defined only by the reversibility of the logical states, which is related only to the logical entropy.  In contrast, the thermodynamic reversibility is related to  the reversibility of the relevant total system (i.e., the whole universe) including the heat bath,  and to the total entropy production as discussed in Sec.~2.
Therefore, these two reversibilities are not equivalent in general.
We note that, if the erasure is not quasi-static but is performed with a finite velocity, the erasure becomes thermodynamically irreversible.  We summarize the thermodynamic and logical reversibilities for the information erasure in Table 1.

\begin{table}
\caption{Summary for the conventional setup of information erasure.}
\begin{center}
\begin{tabular}{| l || l | l | }\hline
{} & Quasi-static & Finite-velocity  \\ \hline \hline
Thermodynamically & reversible & irreversible \\ \hline
Logically & irreversible & irreversible \\ \hline
Heat emission & $ = \beta^{-1} \ln 2 $ & $> \beta^{-1} \ln 2 $ \\ \hline
\end{tabular}
\end{center}
\end{table}

Let us discuss the above point in more detail.  During the information erasure, the entropy is transferred from the logical (i.e., accessible) degrees of freedom in the memory into the environmental (i.e., inaccessible or microscopic) degrees of freedom in the bath.  
Even if the erased information may still be kept in the degrees of freedom of the environment in principle, one cannot  access or recognize in practice such dissipated information in the bath in a real computation.
This implies that the erasure is logically irreversible (i.e., the accessibility to the stored information is lost).
We note that  the logical reversibility depends on the boundary between the logical and environmental degrees of freedom.
For example, in terms of the reversible extension of the erasure~(\ref{ERASE_ex}), the erased information is kept in ancilla  $\mathcal M''$. 
If all of the microscopic physical states in the whole universe were regarded as ``logical states,'' the information erasure could be logically reversible.

In contrast, the above-mentioned entropy transfer from the accessible to inaccessible degrees of freedom does not imply the thermodynamic irreversibility.  
The second law of thermodynamics is relevant only to the reversibility and irreversibility of the whole universe from the macroscopic point of view, as discussed in Sec.~II.
We stress that the thermodynamic reversibility does not depend  on the boundary between the logical and environmental degrees of freedom.
This shows the fundamental difference between the logical and thermodynamic reversibilities.

\

\begin{figure}[htbp]
\begin{center}
  \includegraphics[width=100mm]{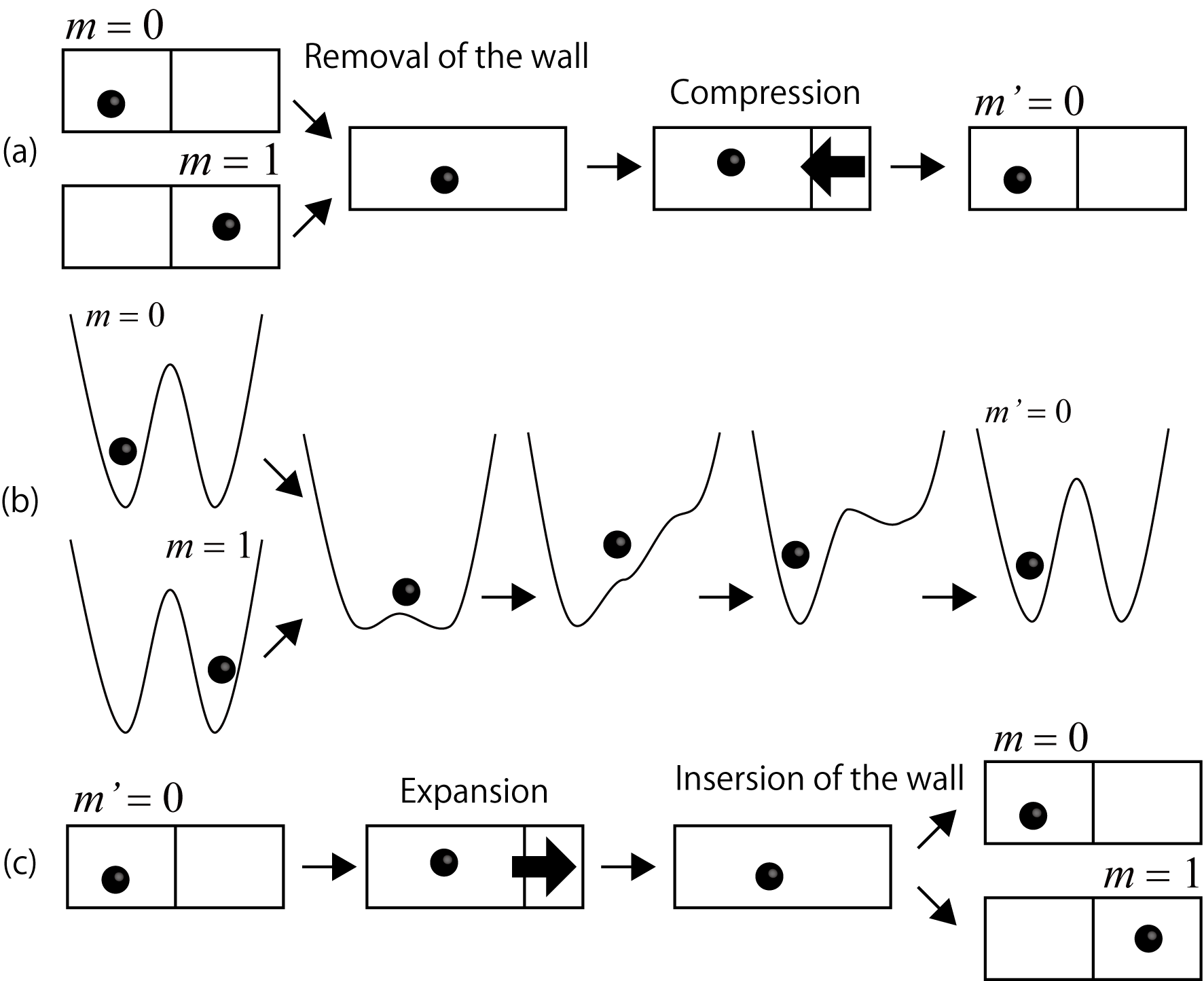}
 \end{center}
\caption{
(a) Information erasure with the two-box memory (Fig.~1 (b)).  The logical state of the memory is initially $0$ or $1$ with equal probability $1/2$, which corresponds to one bit of information.  To erase information, we remove the wall, and compress the box from right quasi-statically and isothermally.  The logical state of the memory then becomes $0$ with unit probability.  While the information erasure is logically irreversible, this  quasi-static erasure protocol achieves the Landauer bound~(\ref{Landauer1}),  and is thermodynamically reversible.
(b) Information erasure in the memory with the binary symmetric potential.  The erasure in (a) with the two-box memory is an idealization of this erasure.
(c) The time-reversal of the information erasure.  The logical state is initially $0$ with unit probability. It finally becomes $0$ or $1$ with equal probability $1/2$, where the probability distribution is the same as that before the erasure.  This process is the time-reversal of (a) in terms of the time evolution of the probability distribution.}
\end{figure}

A simple model of the information erasure that achieves the Landauer bound is illustrated as follows. 
We consider the information erasure with  the two-box memory shown in Fig.~2 (a), which is an idealization of the information erasure with the binary-potential model as shown in Fig.~2 (b).   
Let $V/2$ be the volume of each box.
Before the erasure, the particle is in the left or right box with equal probability $1/2$.  We then remove the wall without performing any work on the memory, and next compress the box from right to left quasi-statically and isothermally.  The particle is then in the left box with unit probability.    The work that is performed on the memory during the compression process is given by
\begin{equation}
W = \int_{V/2}^V \frac{\beta^{-1}}{V'} dV' = \beta^{-1} \ln 2,
\end{equation}
where we used the equation of states of the ideal gas with a single particle.
Therefore, the Landauer bound~(\ref{Landauer2}) is achieved in this erasure protocol. 
By noting that $Q = -W$ holds in this case, the equality in (\ref{Landauer1}) is also achieved.

We show  in Fig.~2 (c)  the time-reversal of the foregoing protocol of the information erasure with the two-box memory.  The logical state of the memory is initially $0$ with unit probability, which is the same as that  after the information erasure.  We then expand the left box quasi-statically and isothermally, so that the volume of the box becomes twice.  During this process, we can extract $\beta^{-1} \ln 2$ of work, and $\beta^{-1}\ln 2$ of heat is absorbed by the memory from the bath.   We next insert a wall at the center of the box, and the final logical state of the memory  becomes $0$ or $1$ with equal probability $1/2$.  While the final logical state may be  different from the pre-erasure logical state for the individual  processes, the probability distribution of the final logical states are the same as that of the pre-erasure logical states.  In fact, the protocol shown in Fig.~2 (c) is the time-reversal of that in Fig.~2 (a) in terms of the ensemble; the time evolutions of their probability distributions are the time-reversal with each other.  We stress that the thermodynamic reversibility is defined in terms of the  ensemble.

We discuss how to achieve the quasi-static limit in the information erasure, in particular for the removal process of the barrier.  In the case of the two-box memory shown in Fig.~2 (a), the velocity of the removal of the wall does not affect the probability distribution of the position of the particle, which is the same before and after the removal.  Therefore, we can rapidly remove the wall even in the quasi-static limit. We note that, if the particle is quantum, its wave function can be affected by the velocity of the removal of the wall~\cite{SWKim}. 

On the other hand,  in the case of the binary-potential memory shown in Fig.~2 (b), the velocity of the removal of the barrier affects the probability distribution of the particle, because the barrier is not infinitely thin but has a finite width. 
In this case, one needs infinitely slow change in the height of the barrier to achieve the quasi-static limit.  
The higher the  barrier is, the more time one needs to achieve the quasi-static limit, because the relaxation time over the two wells becomes exponentially larger as the barrier becomes higher.
This makes it hard to achieve the quasi-static limit  with the binary-potential memory in practice.

\section{Thermodynamics of Computation}

In this section, we discuss the general formulation of  thermodynamics of computation, and derive the general formulas that set the fundamental lower bounds of the work requirement and the heat emission during a computation.  Moreover, we clarify the relationship between the thermodynamic and logical reversibilities in the general setup. 

\subsection{General Setup}

We first consider the physical structure of the memory.  In general,  the logical states do not have one-to-one correspondence to the physical states of the memory;  there may be a lot of microscopic physical states that correspond to a single logical state.

Let $\mathcal Y$ be the phase space of the memory, where each phase-space point $y \in \mathcal Y$ describes a microscopic physical state of  the memory, and let $\mathcal M$ be the set of the possible  input logical states.
To relate the physical states to logical ones, we decompose $\mathcal Y$ into subspaces $\mathcal Y_m$'s ($m \in \mathcal M$), where $\mathcal Y_m$ and $\mathcal Y_n $ do not overlap with each other for $m \neq n$, i.e., $\cup_{m \in \mathcal M} \mathcal Y_m = \mathcal Y$ and $\mathcal Y_m \cap \mathcal Y_n = \phi$ ($m \neq n$) with $\phi$ the empty set.  If the phase-space point of the memory is in $\mathcal Y_m$ before the computation, we define that the input logical sate is $m$.    We call $\mathcal Y_m$ an input logical subspace associated with $m$.

Figure~3 shows simple examples of phase-space separations with binary potentials of the memory.   Figure~3 (a) shows a symmetric potential,  which is the same as the memory illustrated in Fig.~1 (a).  In this case, $0$ and $1$ correspond to the left and right well, respectively, and the phase-space volume  of $\mathcal Y_0$ is the same as that of $\mathcal Y_1$.  
Figure~3 (b) shows an asymmetric potential, where $0$ and $1$ correspond to the left and right well, respectively.  In this case, the phase-space volume  of $\mathcal Y_0$ is different from that of $\mathcal Y_1$.

\begin{figure}[htbp]
\begin{center}
\includegraphics[width=75mm]{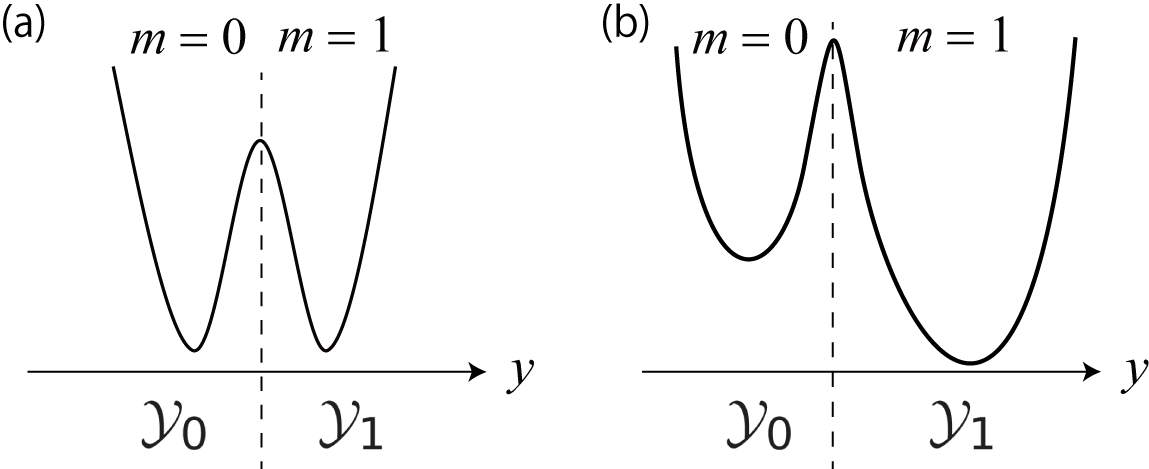}
\end{center}
\caption{
(a) Separation of phase-space $\mathcal Y$ into subspaces $\mathcal Y_0$ and $\mathcal Y_1$ with equal volumes, which respectively correspond to the left and right well of the binary symmetric potential.
(b) Separation of phase-space $\mathcal Y$ into subspaces $\mathcal Y_0$ and $\mathcal Y_1$ with different volumes, which respectively correspond to the left and right well of the asymmetric potential.}
\end{figure}

We consider probability distributions of the physical states in $\mathcal Y$.
Let $y \in \mathcal Y$ be the initial phase-space point before the computation, $P  [y]$ be its  probability,  $m \in \mathcal M$ be the initial logical state such that $y \in \mathcal Y_m$, and $P  [m]$ be its probability that satisfies
\begin{equation}
P[m] = \int_{y \in \mathcal Y_m} dy P[y].
\end{equation}
The probability distribution of $y$ under the condition of $m$ is written as $P  [y|m]$, which takes nonzero value only if $y \in \mathcal Y_m$.  We then have
\begin{equation}
P [y] = \sum_{m \in \mathcal M} P [y|m] P [m].
\end{equation}

We also consider output logical states and phase-space points after the computation.  Let  $\mathcal M'$ be the set of the output logical states, and $\mathcal Y'_{m'}$  be the logical subspace  associated with $m' \in \mathcal M'$, where $\cup_{m' \in \mathcal M'} \mathcal Y'_{m'} = \mathcal Y$ and $\mathcal Y'_{m'} \cap \mathcal Y'_{n'} = \phi$ ($m' \neq n'$).
Let $y' \in \mathcal Y$ be the final phase-space point after the computation,  $P' [y']$ be its  probability,  $m' \in \mathcal M'$ be the final logical state such that $y' \in \mathcal Y'_{m'}$, and $P' [m']$ be its probability.
The  probability  of $y'$ under the condition of $m'$ is written as $P' [y'|m']$, which satisfies
\begin{equation}
P'[y'] = \sum_{m' \in \mathcal M'} P' [y'|m'] P'[m'].
\end{equation}

\subsection{Entropy Balance}

We next consider the changes in entropies during the computation.
Before the computation, the Shannon entropy of the physical states is given by
\begin{equation}
S  (\mathcal Y ) = -\int_{y \in \mathcal Y} dy P [y ] \ln P  [ y ],
\end{equation}
and the logical entropy of the  input  is given by
\begin{equation}
H  (\mathcal M) = -\sum_{m \in \mathcal M} P [m] \ln P  [m].
\end{equation}
The conditional Shannon entropy inside $\mathcal Y_m$ is given by
\begin{equation}
S  (\mathcal Y | m) = -\int_{y \in \mathcal Y_m} dy P [y | m] \ln P  [ y| m],
\end{equation}
whose ensemble average over $m \in \mathcal M$ is
\begin{equation}
S  (\mathcal Y | \mathcal M) := \sum_m P  [m]  S  (\mathcal Y | m).
\end{equation}
Due to a general formula in probability theory~\cite{Cover-Thomas}, these entropies  satisfy 
\begin{equation}
S  (\mathcal Y) = H  (\mathcal M) + S  (\mathcal Y | \mathcal M),
\label{entropy_balance1}
\end{equation}
which implies that  the total entropy  is given by the sum of the logical entropy of $\mathcal M$ and the average of the conditional entropies of $\mathcal Y_m$'s.
Intuitively, the fluctuation over the  whole phase space can be decomposed into that over the logical states and that over the internal physical states in the individual logical subspaces.

We also consider the entropies after the computation in the same manner:
\begin{equation}
H' (\mathcal M') = -\sum_{m' \in \mathcal M'} P' [m'] \ln P' [m'],
\end{equation}
\begin{equation}
S' (\mathcal Y ) = -\int_{y' \in \mathcal Y} dy' P'[y' ] \ln P' [ y' ].
\end{equation}
\begin{equation}
S' (\mathcal Y | m') = -\int_{y' \in \mathcal Y'_{m'}} dy' P' [y' | m'] \ln P' [ y'| m'],
\end{equation}
and
\begin{equation}
S' (\mathcal Y | \mathcal M') := \sum_{m'} P' [m']  S' (\mathcal Y | m').
\end{equation}
They also satisfy the decomposition formula:
\begin{equation}
S' (\mathcal Y) = H' (\mathcal M') + S' (\mathcal Y | \mathcal M').
\label{entropy_balance2}
\end{equation}

Therefore, the total entropy change during the computation is decomposed as 
\begin{equation}
\Delta S = \Delta H  + \Delta S_{\rm in},
\label{total_entropy_change}
\end{equation}
where
\begin{equation}
\Delta S := S' (\mathcal Y) - S(\mathcal Y), 
\label{entropy_change1}
\end{equation}
\begin{equation}
\Delta H := H'(\mathcal M') - H(\mathcal M),
\end{equation}
\begin{equation}
\Delta S_{\rm in} := S' (\mathcal Y | \mathcal M') - S(\mathcal Y | \mathcal M).
\end{equation}
Here, $\Delta S$ is the change in the total entropy, $\Delta H$ is the change in the logical entropy, and $\Delta S_{\rm in}$ is the average of the change in the physical entropy in the individual logical subspaces.
We note that $\Delta H \leq 0$ holds for  logically irreversible computations, while $\Delta H = 0$ holds for  logically reversible computations.

As a special case, we consider the information erasure.
We refer to one of the logical states in $\mathcal M'$ as the ``standard state,'' which we denote by $0 \in \mathcal M'$.
The information erasure is defined as the process in which the output logical state is in the standard state with unit probability (i.e., $P'[0] = 1$ and $P'[m'] = 0$ if $m' \neq 0$) for any probability distribution $P[m]$ of the input  logical states.
In this case, $H' (\mathcal M') = 0$ holds by definition, and therefore $\Delta  H = - H(\mathcal M')$.  The change in the physical entropy is given by
\begin{equation}
\Delta S_{\rm in} = S' (\mathcal Y' |  0) - S(\mathcal Y | \mathcal M).
\end{equation}
We note that  $\Delta S \neq - H(\mathcal M')$ if $\Delta S_{\rm in} \neq 0$.

To clarify the role of $\Delta S_{\rm in}$, we consider a simple case with $\mathcal M = \mathcal M'$ and $\mathcal Y_m = \mathcal Y'_m$.  We assume that the initial and final distributions inside the individual logical subspaces are the same (i.e., $P[y|m] = P'[y|m]$ for any $y$ and $m$), and therefore $S(\mathcal Y| m) = S'(\mathcal Y|m)$ holds for any $m$.  On the other hand, the probability distribution over $\mathcal M$ changes from $P[m]$ to $P'[m]$ during the computation.  In this case, we have
\begin{equation}
\Delta S_{\rm in} = \sum_{m \in \mathcal M} (P'[m] - P[m]) S(\mathcal Y | m).
\end{equation}

If $S(\mathcal Y | m) $ does not depend on $m$, we have $\Delta S_{\rm in} = 0$ for any $P[m]$ and $P'[m]$. 
For example, in the case of the symmetric memory in Fig.~3 (a), $S(\mathcal Y | 0)  = S(\mathcal Y | 1) $ holds if the conditional probability distributions in the individual wells are in thermal equilibrium.   If $\Delta S_{\rm in} = 0$, the internal fluctuations inside the individual logical subspaces do not contribute to the  change in the total entropy (i.e., $\Delta S = \Delta H$).

In contrast, if $S(\mathcal Y | m) $ depends on $m$,  $\Delta S_{\rm in} \neq 0$ in general.  
For example, in the case of the asymmetric memory in Fig.~3 (b), $S(\mathcal Y | 0) $ and $S(\mathcal Y | 1) $ are different with each other.  
If $\Delta S_{\rm in} \neq 0$,  the internal fluctuations inside the individual wells contribute to the change in the total entropy, and therefore $\Delta S \neq \Delta H$.
The role of the asymmetry of the potential has been discussed in Refs.~\cite{Barkeshli,Norton,Maroney,Turgut,Sagawa-Ueda2}.

\subsection{Generalized Landauer Principle}

We now consider the  second law of thermodynamics for computation.
We assume that the memory is attached to a single heat bath at inverse temperature $\beta$ during the computation.
The total entropy production is given by Eq.~(\ref{total_entropy_production}), which leads to
\begin{equation}
\Delta S_{\rm tot} =  \Delta H  + \Delta S_{\rm in} - \beta Q,
\label{total_entropy1}
\end{equation}
where  $Q$ is the average heat that is absorbed by the memory from the bath.  Therefore, the second law~(\ref{second1_1}) is given by
\begin{equation}
\Delta H  \geq \beta Q -  \Delta S_{\rm in},
\label{g_Landauer1}
\end{equation}
or equivalently,
\begin{equation}
-\beta Q \geq - \Delta H  - \Delta S_{\rm in},
\label{g_Landauer2}
\end{equation}
which gives the fundamental lower bound of the heat emission into the bath during the computation.
We refer to inequalities~(\ref{g_Landauer1}) and (\ref{g_Landauer2}) as the generalized Landauer principle, and the right-hand side of (\ref{g_Landauer2}) as the generalized Landauer bound.
Several inequalities that are similar to or essentially equivalent to (\ref{g_Landauer2}) have been obtained in Refs.~\cite{Maroney,Turgut,Sagawa-Ueda2}.

In the special case of  $\Delta S_{\rm in} = 0$, inequality~(\ref{g_Landauer2}) reduces to
\begin{equation}
-\beta Q \geq - \Delta H.
\end{equation}
In this case, if the computation is logically irreversible, the heat emission $-Q$ is nonnegative because of $\Delta H \geq 0$.

In the case of the information erasure, the heat emission is bounded as
\begin{equation}
-\beta Q \geq H  (\mathcal M)  - \Delta S_{\rm in},
\label{g_Landauer3}
\end{equation}
where $\Delta S_{\rm in}$ is the modification term to the original Landauer principle  due to the change in the fluctuations inside the individual logical subspaces.
If $\Delta S_{\rm in} = 0$, we obtain
\begin{equation}
-\beta Q \geq H  (\mathcal M),
\end{equation}
which is the conventional Landauer principle~\cite{Landauer,Shizume,Piechocinska,Esposito2}.

In the conventional setup of the Landauer principle with a symmetric memory  such as Fig.~1 (a) and Fig.~3 (a), the decrease in the entropy of the logical states should be compensated for by the increase in the entropy of the bath, which is accompanied by at least $\beta^{-1} H (\mathcal M)$ of heat emission into the bath.  
In contrast, in the case of the asymmetric memories such as Fig.~3 (b), the decrease in the entropy of the logical states can be compensated for not only by the increase in the entropy of the bath, but also by that inside the individual logical subspaces.
Here, the heat emission is determined only by the change in the entropy of the bath, but not by that inside the logical subspaces.
Therefore, the lower bound of the heat emission can be different from  $\beta^{-1}  H (\mathcal M)$ as shown in inequality~(\ref{g_Landauer3}); in particular, the heat emission can be smaller than  $\beta^{-1}  H (\mathcal M)$  if $\Delta S_{\rm in} > 0$.

To illustrate the above situation, we consider the situation with $\mathcal M = \mathcal M' = \{ 0, 1 \}$ and $\mathcal Y_m = \mathcal Y'_m$ for $m=0,1$. We assume that $S (\mathcal Y | 0) = S' (\mathcal Y |  0 )$ holds, which implies that the probability distribution inside the standard state is the same before and after the erasure.  We then have
\begin{equation}
\eqalign{
\Delta S_{\rm in} &= S (\mathcal Y | 0) - \left( P[0] S (\mathcal Y | 0) + P[1] S (\mathcal Y | 1) \right) \\
&= P[1] \left( S (\mathcal Y | 0) -  S (\mathcal Y | 1) \right).
}
\end{equation}
Therefore, if $P [1] \neq  0$ and $S (\mathcal Y | 0) >  S (\mathcal Y | 1)$, we have  $\Delta S_{\rm in} > 0$. 
It is natural to assume that $S (\mathcal Y | 0) \neq  S (\mathcal Y | 1)$ in asymmetric memories.

\subsection{Thermodynamic and logical reversibilities}

We now  summarize the relationship between the thermodynamic and logical reversibilities in the general setup. 
On the basis of the argument in Sec.~5.2,  the conventional Landauer principle discussed in Sec.~4 needs to be modified in general, and the generalized Landauer principle is  stated as follows: \\
\\
{\it The decrease in the entropy of the logical states during a logically irreversible computation can be compensated for not only by the increase in the  entropy  of  the heat bath, but also by that of the physical states inside the individual logical subspaces, where only the former determines the amount of the heat emission.}\\
\

Since the generalized Landauer principle (\ref{g_Landauer1}) is equivalent to the second law of thermodynamics~(\ref{second1}), the equality in (\ref{g_Landauer1}) can be achieved in the quasi-static limit (i.e., in the case of the quasi-static computation) where the total entropy production is zero.  In this case, the computational process is thermodynamically reversible.  Therefore, we conclude that:\\
\\
{\it Any logically irreversible computation can be performed in a thermodynamically reversible manner in the quasi-static limit, and therefore, the thermodynamic and logical reversibilities are not equivalent with each other.}\\
\\
These are generalizations of the arguments in  Sec.~4.  We again stress that our observations here do not contradict the conventional Landauer principle;   the logical reversibility is related to the change in the entropy of the logical states, while the thermodynamic reversibility is related to the change in the entropy of the whole universe that consists of the logical and physical states of the memory and the physical states of the heat bath.


\subsection{Work requirement for computation}

We next consider the thermodynamic work needed for computation.
Let $E [y]$ with $y \in \mathcal Y$ be the initial Hamiltonian of the memory.  We define the conditional free energy in  logical subspace $\mathcal Y_m$ as
\begin{equation}
F_m := -\beta^{-1} \int_{y \in \mathcal Y_m} dy e^{-\beta E[y]}.
\end{equation}
We assume that the memory is initially  in the conditional canonical distribution in the individual logical subspaces, which is given by
\begin{equation}
P [y | m ] = e^{\beta (F_m - E [y])}
\end{equation}
for $y \in \mathcal Y_m$, and otherwise $P [y | m ] = 0$.  The total probability distribution is then given by
\begin{equation}
P [y] = \sum_{m \in \mathcal M} P[m] \chi (y, m) e^{\beta (F_m - E [y])},
\end{equation}
where $\chi (y, m)$ is the characteristic function which takes one if $y \in \mathcal Y_m$ and  zero otherwise.
The conditional entropy is then given by
\begin{equation}
S(\mathcal Y | m) = \beta (E_m - F_m), 
\end{equation}
where 
\begin{equation}
E_m := \int_{y \in \mathcal Y_m} dy P[y|m] E[y].
\end{equation}
Therefore, we obtain
\begin{equation}
S(\mathcal Y | \mathcal M) = \beta ( E -  F ),
\end{equation}
where
\begin{equation}
E  := \sum_{m \in \mathcal M} P[m] E_m = \int_{y \in \mathcal Y} dy P[y] E[y]
\end{equation}
is the average energy, and 
\begin{equation}
F  := \sum_{m \in \mathcal M} P[m] F_m
\end{equation}
is the average free energy.

We also consider the final Hamiltonian $E'[y']$ after the computation.  Correspondingly, every argument about the final state is parallel to that about the initial one in the previous paragraph.  To show what a quantity is about the final states after the computation, we use notation of prime.


We then have
\begin{equation}
\Delta S_{\rm in} = \beta (  \Delta E  -  \Delta F ),
\label{Sint_F}
\end{equation}
where $\Delta E :=  E' - E $ is the change in the average energy, and $\Delta F :=  F'  - F$ is the change in the average free energy.
We note that, if the final distribution is different from the conditional canonical distribution, we can show an inequality:
\begin{equation}
\Delta S_{\rm in} \leq \beta (  \Delta E - \Delta F ),
\label{entropy}
\end{equation}
where the equality in (\ref{entropy}) is achieved if and only if the output state is in the conditional canonical distribution.
Therefore, the total entropy production~(\ref{total_entropy1}) satisfies
\begin{equation}
\Delta S_{\rm tot}  \leq \beta ( W -  \Delta F)  + \Delta H,
\label{total_entropy2}
\end{equation}
where $W$ is the work performed on the memory, and we used the first law of thermodynamics
\begin{equation}
\Delta E = Q + W.
\end{equation}
Therefore, by applying the second law~(\ref{second1}), we obtain the generalized Landauer principle in terms of the work:
\begin{equation}
\beta W \geq  - \Delta H +  \beta \Delta F, 
\label{g_LandauerW1}
\end{equation}
which gives the minimal work needed for the computation.  
In the case of information erasure, inequality~(\ref{g_LandauerW1}) reduces to
\begin{equation}
\beta W_{\rm eras} \geq  H (\mathcal M) + \beta \Delta F,
\label{g_Landauer_work}
\end{equation}
where $\Delta F$  is the modification term to the conventional Landauer principle~(\ref{Landauer2}). As shown in Eq.~(\ref{Sint_F}), there are two contributions to $\Delta F$: the changes in the entropy and the energy inside the individual logical subspaces, where the former is $\Delta S_{\rm in}$ and the latter is  $\Delta E$. 
We can also regard $W - \Delta F$ as the energy cost needed for the information erasure, whose lower bound is given by the conventional Landauer bound $\beta^{-1} H(\mathcal M)$. 
Several inequalities that are similar to or essentially equivalent to the generalized Landauer principle (\ref{g_Landauer_work}) have been obtained in Refs.~\cite{Maroney,Turgut,Sagawa-Ueda2}.

In general, we call the memory as symmetric if  $F_m$ does not depend on $m$.   The memory shown in Fig.~3 (a) is  symmetric  in this sense because of $F_0 =F_1$.  In contrast, the memory shown in Fig.~3 (b) is asymmetric because of $F_0 \neq F_1$.
When $\mathcal M = \mathcal M'$, $\mathcal Y_m = \mathcal Y'_{m'}$ with $m=m'$,  $F_m = F'_{m'}$ with $m=m'$, and $F_m$ does not depend on $m$,  then we have $\Delta F=0$ for any  $P[m]$ and $P'[m]$.

As a simple example of information erasure with an asymmetric memory, we consider a model of the memory shown in Fig.~4 (a) that is in contact with a heat bath at inverse temperature $\beta$.  If the barrier is much higher than the thermal fluctuation, this model is idealized by a memory with two boxes with different volumes as shown in Fig.~4 (b).  Let $t:1-t$ ($0 < t < 1$) be the ratio of the volumes of the boxes, which characterizes the ratio of the phase-space volumes inside the individual logical states.
If the memory is symmetric, $t = 1/2$.  
We assume that $\mathcal M = \mathcal M' = \{ 0,1 \}$ and $\mathcal Y_m = \mathcal Y'_m$ with $m=0,1$, and write $p := P[0]$.  We note that $t$ is in general different from $p$, because the initial state is not necessarily in global thermal equilibrium over the whole phase space.
The protocol of the information erasure is as follows (see also Fig.~4 (c)).  We first move the wall quasi-statically to the position where the ratio of the two volumes is given by $p:1-p$.   
During this process, we perform the work of  $\beta^{-1} [ p \ln (t/p) + (1-p) \ln (1-t/1-p)]$ on average.  We then remove the wall without any work.  We next compress the box from the right quasi-statically, and the ratio of the two volumes is given by $t:1-t$ in the final stage, where the logical state is ``$0$'' with unit probability. During this compression, we perform the work of $-\beta^{-1}\ln t$.  The total work performed on the memory is then given by
\begin{equation}
\beta W_{\rm eras} = - p \ln p - (1-p) \ln (1-p) + (1-p \ln \frac{1-t}{t}).
\label{erasure_asymmetric}
\end{equation}
We note that the initial logical entropy is given by $H(\mathcal M) = - p \ln p - (1-p ) \ln (1-p)$.  The free-energy difference between two logical states are given by  $F_0 - F_1 = \beta^{-1} \ln [(1-t)/t]$, and therefore $\Delta F = \beta^{-1} (1-p) \ln [(1-t)/t]$.  Therefore, Eq.~(\ref{erasure_asymmetric}) achieves the generalized Landauer bound in~(\ref{g_Landauer_work}), which implies that this protocol of the information erasure is thermodynamically reversible.  

\

\begin{figure}[htbp]
\begin{center}
\includegraphics[width=100mm]{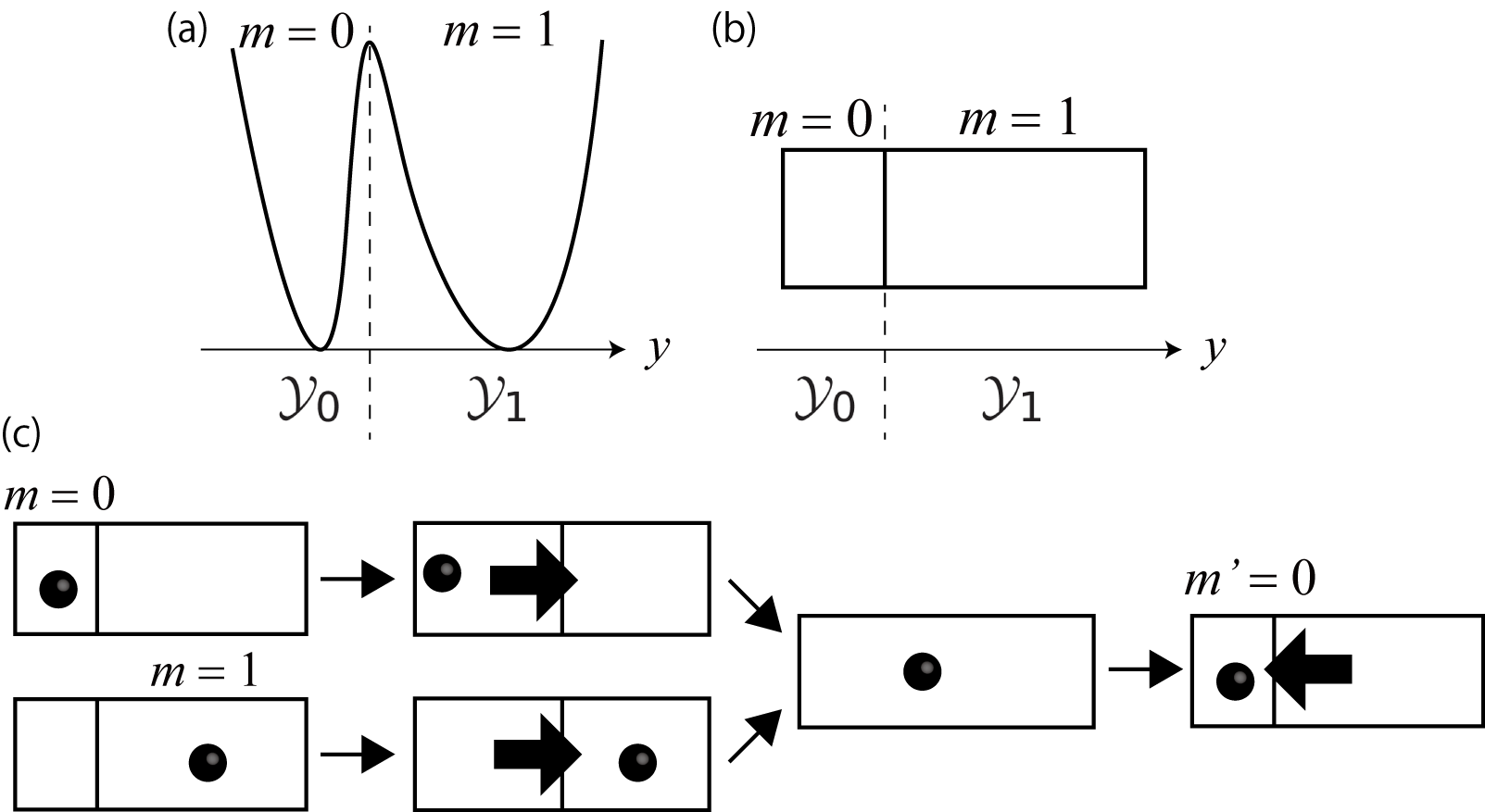}
\end{center}
\caption{
(a) An asymmetric memory with phase-space separation $\mathcal Y_0$ and $\mathcal Y_1$ with different volumes.  
(b) An asymmetric memory with two boxes with different volumes, which is an idealization of the asymmetric memory in (a).
(c) Information erasure with the asymmetric memory (b).  The logical state of the memory is initially $0$ or $1$ with equal probability $1/2$, which corresponds to one bit of information.  We move the wall quasi-statically and isothermally to the center of the box,  remove the wall, and compress the box from the right quasi-statically and isothermally so that the final logical state is $0$ with unit probability.  This erasure protocol achieves the generalized Landauer bound in~(\ref{g_Landauer_work}), and therefore is thermodynamically reversible.}
\end{figure}

We next consider a simple example of the information erasure with $\mathcal M \neq \mathcal M'$.  We assume that $\mathcal M = \{ 0,1 \}$ and that the memory is initially binary symmetric as shown in Fig.~1 (a), which is idealized by the two-boxes memory with equal volumes as shown in Fig.~1 (b).  We then assume that $\mathcal M' = \{ 0 \}$ in the final stage, i.e., the output logical state is only the standard state; the corresponding potential model  is shown in Fig.~5 (a).   This is idealized by  a memory with a single box as shown in Fig.~5 (b).  For simplicity, we set $P[0] = 1/2$.  In this case, the information erasure is just the removal of the wall as shown in Fig.~5 (c), and therefore $W_{\rm eras} = 0$.   This erasure protocol achieves the generalized Landauer bound in~(\ref{g_Landauer_work}), since  $H(\mathcal M) = \ln 2$, $F_0 - F'_0 = F_1 - F'_0 = \beta^{-1} \ln 2$, and $\Delta F = - \beta^{-1} \ln 2$.  Therefore, this erasure protocol is thermodynamically reversible.  In fact, if we insert a wall to the center of the single box after the erasure, the probability distribution of the logical states  returns to the initial one, where a particle is in one of the two boxes with equal probability $1/2$.

\begin{figure}[htbp]
\begin{center} 
\includegraphics[width=100mm]{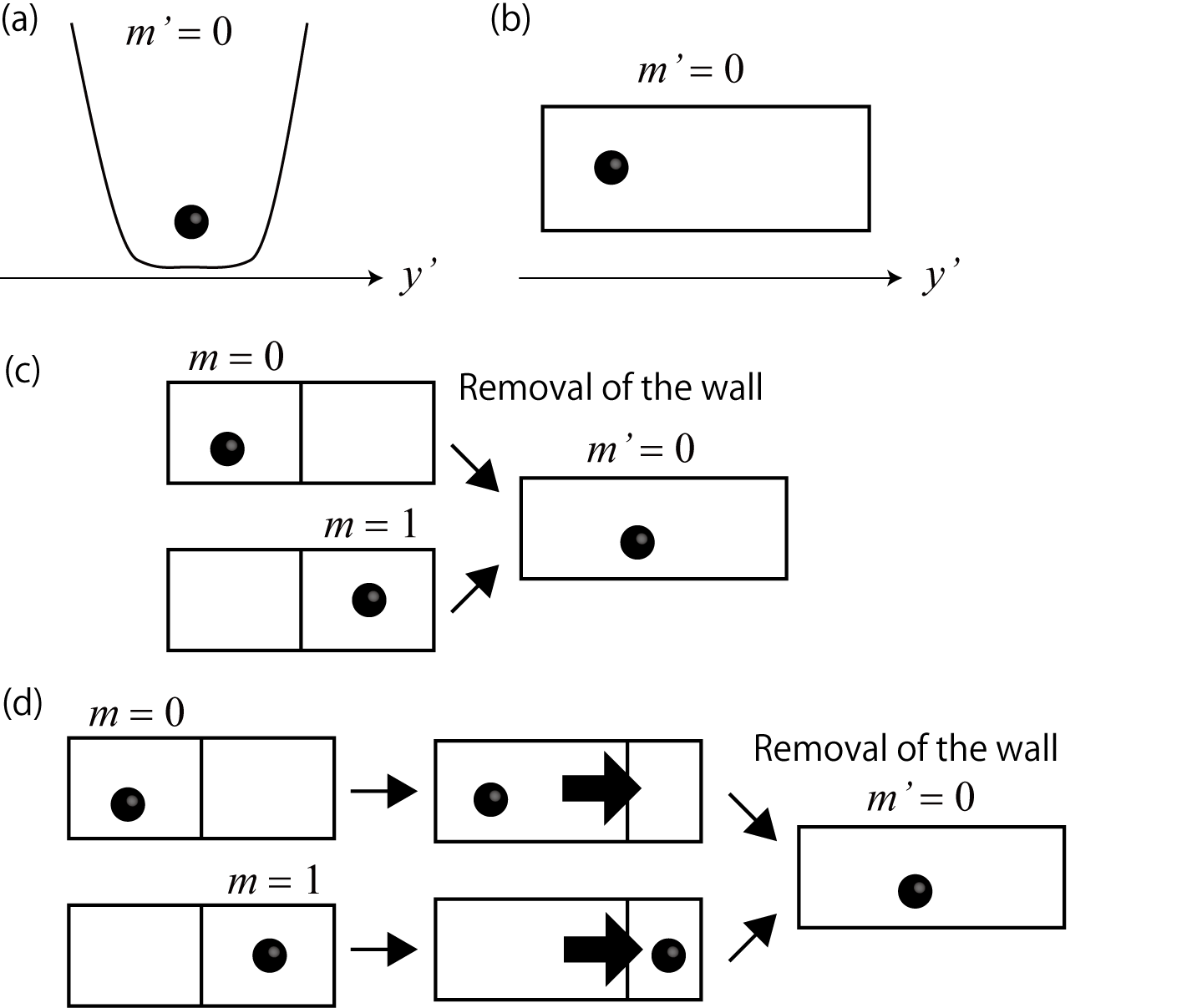}
\end{center}
\caption{
(a) A memory with a single-well potential, whose logical state is only the standard state $m'=0$.
(b) A memory with a single box, which is  an idealization of the single-well memory in (a).
(c) Information erasure from the symmetric two-box memory [Fig.~1 (b)] to the single-box memory (b).  The initial logical state of the memory is $m=0$ or $1$ with equal probability $1/2$, which corresponds to one bit of information.   We then only remove the wall so that the final logical state becomes $m'=0$, which can be regarded as the information erasure that achieves the generalized Landauer bound in~(\ref{g_Landauer_work}).
(d)  Information erasure where the initial logical state is $0$ ($1$) with probability $p$ ($1-p$). We first move the wall  quasi-statically and isothermally so that the ratio of the volumes of the boxes becomes $p:1-p$.  We then remove the wall, and the final logical state becomes $m'=0$ with unit probability.  This erasure also achieves the generalized Landauer bound in~(\ref{g_Landauer_work}).}
\end{figure}

If $p := P[0] \neq 1/2$ in the above model, we  quasi-statically move the wall to the position where the ratio of the volumes $p:1-p$, and then remove the wall (see all Fig.~5 (d)).  In this case, the work performed on the memory is given by $\beta W_{\rm eras} = - \ln 2 - p \ln p - (1-p) \ln (1-p)$.  Since $H(\mathcal M) = - p \ln p - (1-p) \ln (1-p)$ and $\beta \Delta F = - \ln 2$, this erasure protocol also achieves the generalized Landauer bound in~(\ref{g_Landauer_work}).

\section{Two Memories}

In this section, we consider the cases that  the memory consists of  two sub-memories.  In particular, we discuss measurement and feedback control between the two sub-memories, which constitute the typical setup of Maxwell's demon~\cite{Demon}.

\subsection{General argument}

We assume that any logical state in  $\mathcal M$ is a pair of two logical states of sub-memories that we refer to as S and D; any input logical state $m \in \mathcal M$ is written as $m = (s,d)$ where $s$ and $d$ are the logical states of memories S and D, respectively.
Let $\mathcal M^{\rm S}$ ($\mathcal M^{\rm D}$) be the set of logical states of S (D) with  $\mathcal M = \mathcal M^{\rm S} \times \mathcal M^{\rm D}$, where  $s \in \mathcal M^{\rm S}$ and $d \in \mathcal M^{\rm D}$.

We also consider the physical states of the memories.  Let $\mathcal Y^{\rm S}$ ($\mathcal Y^{\rm D}$) be the set of physical states of S (D) with  $\mathcal Y = \mathcal Y^{\rm S} \times \mathcal Y^{\rm D}$.
Let   $y := (u,v)$ for any initial physical state $y \in \mathcal Y$ with $u \in \mathcal Y^{\rm S}$ and $v \in \mathcal Y^{\rm D}$.  
Let $\mathcal Y_m := \mathcal Y_s^{\rm S} \times \mathcal Y_d^{\rm D}$.

The conditional probability of $(u,v)$ under the condition of $(s,d)$ is given by $P[u,v | s,d]$ that takes nonzero value only if $u \in \mathcal Y_s^{\rm S}$ and $v \in \mathcal Y_d^{\rm D}$.  By applying the argument in Sec.~5.1 to the present situation, we have
\begin{equation}
P[u,v] = \sum_{(s,d) \in \mathcal M} P[u,v | s,d] P[s,d],
\end{equation}
whose Shannon entropy  is given by Eq.~(\ref{entropy_balance1}).

The mutual information plays a crucial role in the presence of two memories; it characterizes the correlation between them~\cite{Shannon,Cover-Thomas}.  The mutual information between the physical states and that between the logical states are respectively given by
\begin{equation}
I (\mathcal Y^{\rm S} : \mathcal Y^{\rm D}) = S (\mathcal Y^{\rm S}) + S(\mathcal Y^{\rm D}) - S(\mathcal Y),
\end{equation}
\begin{equation}
I (\mathcal M^{\rm S} : \mathcal M^{\rm D}) = H (\mathcal M^{\rm S}) + H(\mathcal M^{\rm D}) - H(\mathcal M).
\end{equation}
The mutual information between the internal states is
\begin{equation}
I (\mathcal Y^{\rm S} : \mathcal Y^{\rm D} | s, d) = S (\mathcal Y^{\rm S} | s) + S(\mathcal Y^{\rm D} | d) - S(\mathcal Y | s,d),
\end{equation}
whose ensemble average over $(s,d)$ is given by
\begin{equation}
\eqalign{
I (\mathcal Y^{\rm S} : \mathcal Y^{\rm D} | \mathcal M) &:= \sum_{(s,d) \in \mathcal M }  I (\mathcal Y^{\rm S} : \mathcal Y^{\rm D} | s, d) 
P[s,d]\\
&= S (\mathcal Y^{\rm S} | \mathcal M^{\rm S}) + S(\mathcal Y^{\rm D} | \mathcal M^{\rm D}) -S(\mathcal Y | \mathcal M).
}
\end{equation}
Therefore, we obtain
\begin{equation}
I (\mathcal Y^{\rm S} : \mathcal Y^{\rm D}) = I (\mathcal M^{\rm S} : \mathcal M^{\rm D})   + I (\mathcal Y^{\rm S} : \mathcal Y^{\rm D} | \mathcal M),
\label{two_decom1}
\end{equation}
which implies that the total correlation between the two memories can be decomposed into the correlation between their logical states and that between their internal physical states in the individual logical subspaces.  We note that $I (\mathcal Y^{\rm S} : \mathcal Y^{\rm D} | \mathcal M) = 0$ if  $P[u,v | s,d] = P [u |s,d] P[v|s,d]$.
We also note that
\begin{equation}
S (\mathcal Y^{\rm S}) = H(\mathcal M^{\rm S}) + S(\mathcal Y^{\rm S} | \mathcal M^{\rm S}),
\label{two_decom2}
\end{equation}
\begin{equation}
S (\mathcal Y^{\rm D}) = H(\mathcal M^{\rm D}) + S(\mathcal Y^{\rm D} | \mathcal M^{\rm D}).
\label{two_decom3}
\end{equation}
The sum of Eqs.~(\ref{two_decom1}), (\ref{two_decom2}), and (\ref{two_decom3}) leads to Eq.~(\ref{entropy_balance1}).

We also consider the probability distributions, the Shannon entropy, and the mutual information after the computation in the parallel manner to those before the computation; for example, we write $\mathcal M' := \mathcal M'_{\rm S} \times \mathcal M'_{\rm D}$,  $m' := (s',d')$ with $s' \in \mathcal M'^{\rm S}$ and $d' \in \mathcal M'^{\rm D}$. To show what a quantity is about the states after computation, we use notation of prime.

The change in the total Shannon entropy is then given by
\begin{equation}
\Delta S = \Delta H^{\rm S} + \Delta H^{\rm D} - \Delta I + \Delta S_{\rm in}^{\rm S} + \Delta S_{\rm in}^{\rm D} - \Delta I_{\rm in}, 
\end{equation}
where
\begin{equation}
\Delta S := S'(\mathcal Y') - S(\mathcal Y),
\end{equation}
\begin{equation}
\Delta H^{\rm S} := H'(\mathcal M'^{\rm S}) - H(\mathcal M^{\rm S}), \ \ \Delta H^{\rm D} := H'(\mathcal M'^{\rm D}) - H(\mathcal M^{\rm D}),
\end{equation}
\begin{equation}
\Delta I := I' (\mathcal M'^{\rm S} : \mathcal M'^{\rm D}) - I (\mathcal M^{\rm S} : \mathcal M^{\rm D}),
\end{equation}
\begin{equation}
\Delta S_{\rm in}^{\rm S} := S'(\mathcal Y'^{\rm S} | \mathcal M'^{\rm S} ) - S(\mathcal Y^{\rm S} | \mathcal M^{\rm S}), \ \ \Delta S_{\rm in}^{\rm D} := S'(\mathcal Y'^{\rm D} | \mathcal M'^{\rm D}) - S(\mathcal Y^{\rm D} | \mathcal M^{\rm D}),
\end{equation}
\begin{equation}
\Delta I_{\rm in} := I' (\mathcal Y'^{\rm S} : \mathcal Y'^{\rm D} | \mathcal M') - I (\mathcal Y^{\rm S} : \mathcal Y^{\rm D} | \mathcal M).
\end{equation}
Here, $\Delta H^{\rm S}$ and $\Delta H^{\rm D}$ describe the changes in the logical entropies, $\Delta I$ describes the change in the logical correlation, $\Delta S_{\rm in}^{\rm S}$ and  $\Delta S_{\rm in}^{\rm D}$ describe the change in the  internal physical entropies in the individual logical subspaces, and $\Delta I_{\rm in}$ describes the change in the  internal correlation between the individual logical subspaces.
Therefore, the total entropy production is given by
\begin{equation}
\Delta S_{\rm tot} = \Delta H^{\rm S} + \Delta H^{\rm D} - \Delta I + \Delta S_{\rm in}^{\rm S} + \Delta S_{\rm in}^{\rm D} - \Delta I_{\rm in}  - \beta Q.
\label{two_total}
\end{equation}

We next assume that the energies of two memories are additive before the computation, $E[y] = E^{\rm S}[u] + E^{\rm D} [v]$, where $E^{\rm S}[u] $ ($E^{\rm D} [v]$) is the Hamiltonian of S (D).  
This assumption implies that the interaction Hamiltonian between the memories is negligible before the computation. 
The corresponding free energies are given by
\begin{equation}
F_s^{\rm S} := - \beta^{-1} \ln \int_{u \in \mathcal Y_s} e^{- \beta E^{\rm S}[u]}, \ \ F_d^{\rm D} := - \beta^{-1} \ln \int_{v \in \mathcal Y_d} e^{- \beta E^{\rm D}[v]}.
\end{equation}
In this case, the conditional distribution is given by
\begin{equation}
P[u,v |s,d] = P[u|s]P[v|d],
\end{equation}
where, if $u \in \mathcal Y_s^{\rm S}$ and $v \in \mathcal Y_d^{\rm S}$,
\begin{equation}
P[u|s] = e^{\beta (F_s^{\rm S} - E^{\rm S} [u])}, \ \ P[v|d] =  e^{\beta (F_d^{\rm D} - E^{\rm D} [v])}.  
\end{equation}
We also consider the energies in the memories after the computation, where the energies are also assumed to be additive.  The argument about energy and the free energy after the computation is parallel to that before the computation.  To show what a quantity is about after the computation, we use notation of prime.  We note that the energies are not necessarily additive during the computation, since the interaction Hamiltonian between the memories may become nonzero  during the computation.

If the initial probability distribution is the conditional probability distribution, the total entropy production satisfies Eq.~(\ref{total_entropy2}).  Therefore, we obtain the lower bound of the work that is needed for the computation with two memories:
\begin{equation}
W \geq - \Delta H  + \Delta F,
\end{equation}
where
\begin{equation}
\Delta H  = \Delta H^{\rm S} + \Delta H^{\rm D} - \Delta I, 
\end{equation}
\begin{equation}
\Delta F := \Delta F^{\rm S} + \Delta F^{\rm D},
\end{equation}
\begin{equation}
\Delta F^{\rm S}  := \sum_{s \in \mathcal M^{\rm S}} P[s] F_s^{\rm S}, \ \  \Delta F^{\rm D} := \sum_{d \in \mathcal M^{\rm D}} P[d] F_d^{\rm D}. 
\end{equation}
In the special case of $\Delta H^{\rm S} = \Delta H^{\rm D} = 0$ and $\Delta F = 0$, we have
\begin{equation}
- W \leq - \Delta I,
\end{equation}
which is the work extraction by using the correlation, which can be demonstrated by Maxwell's demon~\cite{Demon}.

\subsection{Measurement and feedback}

In this subsection,  we consider measurement and feedback as a special case of the argument in Sec.~6.1 with the assumption of the initial and final conditional canonical distributions.  

We first consider a measurement process, where memory D performs a measurement on memory S.  In other words, the information in memory S is copied to memory D. 
The logical state of D is initially the standard state $0^{\rm D}$ with unit probability, and the logical entropy of S is initially given by $H^{\rm S}$.
We assume  that the initial logical entropy in D and the initial mutual information between the logical states are both zero.  
We also assume that the two memories are in the conditional canonical distributions before and after the measurement.

The two memories interact with each other, and make a correlation  between the logical states.
We assume that the logical state of  S does not change in time during the measurement.  
When both of the two memories are one bit, a typical example of such measurement is given by CNOT~(\ref{CNOT_measurement}), where  the measurement is error-free (i.e., the copy of information is perfect).

We denote  by $I$ the correlation between the logical states after the measurement, and denote by $H^{\rm D}$ the logical entropy of D after the measurement.
We note that the mutual information satisfies the following inequalities~\cite{Shannon,Cover-Thomas}:
\begin{equation}
0 \leq I \leq H^{\rm S}, \ \ 0 \leq I \leq H^{\rm D}.
\end{equation}
Applying Eq.~(\ref{total_entropy2}) to the measurement,  the total entropy production is given by
\begin{equation}
\Delta S_{\rm tot} = \beta (W_{\rm meas} - \Delta F_{\rm meas}^{\rm D}) + H^{\rm D} - I,
\label{tot_meas}
\end{equation}
where $W_{\rm meas}$ is the work performed on the memories,  and $\Delta F_{\rm meas}^{\rm D}$ is the change in the average free energy of D.
Therefore, we obtain 
\begin{equation}
W_{\rm meas} \geq   I    - H^{\rm D} + \Delta F_{\rm meas}^{\rm D},
\label{meas_cost1}
\end{equation}
which gives the minimal work needed for the measurement.
Several inequalities that are essentially equivalent to (\ref{meas_cost1}) have been obtained in Refs.~\cite{Sagawa-Ueda2,Sagawa-Ueda2012,Sagawa-Ueda-NJP}.

If $ I  = H^{\rm D}$ holds and  memory D is symmetric (i.e., $\Delta F_{\rm meas}^{\rm D} = 0$), inequality~(\ref{meas_cost1}) reduces to
\begin{equation}
W_{\rm meas} \geq 0,
\label{bound_Bennett}
\end{equation}
which is the bound discussed by Bennett~\cite{Bennett1}.  Since the energies of the memories do not change in time during the measurement if the memory is symmetric, inequality~(\ref{bound_Bennett}) is equivalent to
\begin{equation}
- Q_{\rm meas} \geq 0,
\label{bound_Bennett1}
\end{equation}
where $Q_{\rm meas}$ is the heat absorbed by the memories during the measurement.  Inequalities~(\ref{bound_Bennett}) and (\ref{bound_Bennett1}) imply that the lower bounds of the work requirement and the heat emission during  the measurement are both zero.

We consider a simple example where the logical states of S and D before and after the measurement are all one bit, which is a conventional setup of the measurement~\cite{Bennett1}.
We  assume that S is symmetric (i.e., $F^{\rm S}_0 = F^{\rm S}_1$) before and after the measurement; in this case, memory S  can be modeled by two boxes  (see  also Fig.~6 (a)).
Before the measurement, the logical state $(s,d)$ is $(0,0)$ or $(1,0)$ with equal probability $1/2$.  
The measurement protocol is given by CNOT where S is the control bit and D is the target bit, and then the final logical state is $(0,0)$ or $(1,1)$ with equal probability $1/2$.    In this case, $I = H^{\rm D} = \ln 2$ and $\Delta F_{\rm meas}^{\rm D} = 0$.    Therefore,  $W_{\rm meas} = 0$ and $- Q_{\rm meas} = 0$ are achieved in the quasi-static limit,  where the measurement is thermodynamically reversible as well as logically reversible.   We summarize the thermodynamic and logical reversibilities for the conventional setup of measurement in Table 2, which is contrastive to Table 1 for the information erasure.

\begin{table}
\caption{Summary for the conventional setup of measurement.}
\begin{center}
\begin{tabular}{| l || l | l | }\hline
{} & Quasi-static & Finite-velocity  \\ \hline \hline
Thermodynamically & reversible & irreversible \\ \hline
Logically & reversible & reversible \\ \hline
Heat emission & $ = 0 $ & $> 0 $ \\ \hline
\end{tabular}
\end{center}
\end{table}

\begin{figure}[htbp]
\begin{center}
  \includegraphics[width=100mm]{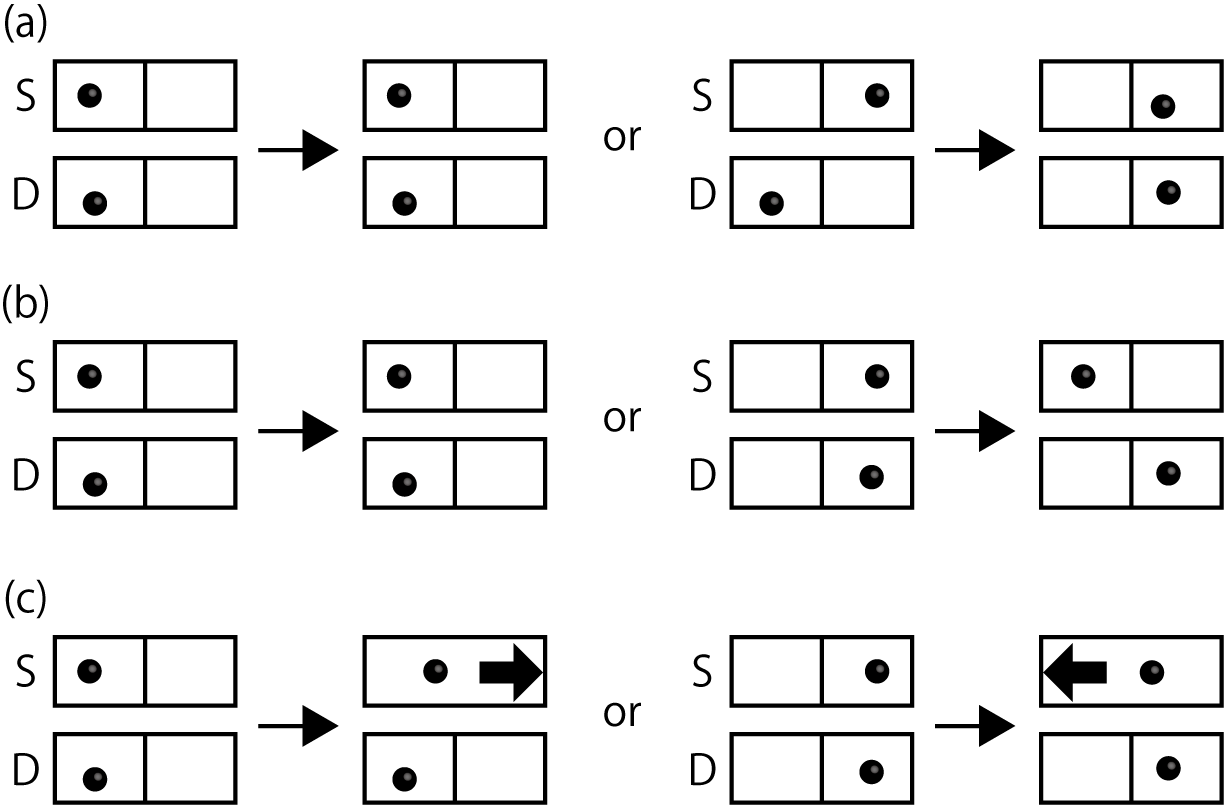}
  \end{center}
\caption{
(a) Model of measurement with two symmetric memories.  The two memories are initially not correlated.  The logical state of D quasi-statically  changes only if the logical state of S is initially in $1$, and the memories finally have one bit of correlation.  This measurement achieves the equality in~(\ref{bound_Bennett}), and is thermodynamically reversible.
(b) Model of feedback control, where the two symmetric memories initially share one bit of correlation.   The logical state of S quasi-statically changes only if the logical state of D is initially in $1$.  Due to the feedback control,  the initial correlation vanishes and the logical entropy of S is decreased by one $\ln 2$.  This feedback protocol achieves the equality in~(\ref{feedback_cost1}), and is thermodynamically reversible.
(c) Model of feedback control, where the two symmetric memories initially share one bit of correlation.  The left (right)  box of S is expanded quasi-statically and isothermally if the initial logical state of D is $0$ ($1$), so that the final logical subspace of S is the whole phase space.  Due to the feedback control, the initial correlation vanishes and $\beta^{-1}\ln 2$ of work is extracted from S, which is equivalent to the case of the Szilard engine.  This feedback protocol also achieves the equality in~(\ref{feedback_cost1}), and is thermodynamically reversible.}
\end{figure}



We next consider a feedback control process followed by the measurement, where memory  D performs feedback control on memory S.
This situation can be regarded as a typical setup of Maxwell's demon (see also Sec.~6.3), where D is the demon and S is the engine to be controlled.
We assume that the two memories are in the conditional canonical distributions before and  after the feedback control.
The logical state of  S changes depending on the logical state of D after the measurement.
We assume that memory  D does not evolve in time during the feedback control.
For simplicity, we assume that the final correlation between the logical states between the memories is zero after the feedback control, such that the change in the logical mutual information is given by $-I$. We also assume that the output logical state of S is the standard state $0^{\rm S}$ with unit probability after feedback control such that the change in the logical entropy of S is given by $-H^{\rm S}$.
When both of the two memories are one bit, a typical example of such feedback control is also given by CNOT~(\ref{CNOT_feedback}).

Applying Eq.~(\ref{total_entropy2}) to the feedback control,  the total entropy production is given by
\begin{equation}
\Delta S_{\rm tot} = \beta (W_{\rm fb} - \Delta F_{\rm fb}^{\rm S}) - H^{\rm S} + I,
\label{tot_feedback}
\end{equation}
where $W_{\rm fb}$ is the work performed on the memories, and $\Delta F_{\rm fb}^{\rm S}$ is the change in the average free energy of S.
Therefore,  we obtain
\begin{equation}
W_{\rm fb} \geq  - I    + H^{\rm S} + \Delta F_{\rm fb}^{\rm S},
\label{feedback_cost1}
\end{equation}
which gives the minimal work needed for the feedback control.
Several inequalities that are essentially equivalent to (\ref{feedback_cost1}) has been obtained in Refs.~\cite{Sagawa-Ueda1,Sagawa-Ueda3,Sagawa-Ueda4,Sagawa-Ueda2012,Sagawa-Ueda-NJP}; in these previous works,  $H^{\rm S} + \Delta F_{\rm fb}^{\rm S}$ is denoted by, for example, just $\Delta F_{\rm fb}^{\rm S}$.

We consider a simple situation where the logical states of S and D before and after feedback are all one bit.  
We assume that S is symmetric (i.e., $F^{\rm S}_0 = F^{\rm S}_1$) before and after the feedback (see also Fig.~6 (b)).
Before feedback, the logical state $(s,d)$ is $(0,0)$ or $(1,1)$ with equal probability $1/2$.  The feedback protocol is given by CNOT that is logically reversible, where D is the control bit and S is the target bit, and then the final logical state is $(0,0)$ or $(0,1)$ with equal probability $1/2$.    In this case, $I = H^{\rm S} = \ln 2$ and $\Delta F_{\rm fb}^{\rm S} = 0$ hold.  Therefore, inequality~(\ref{feedback_cost1}) reduces to $W_{\rm fb} \geq 0$.  In the quasi-static limit, $W_{\rm fb} = 0$ is achieved, which implies that we can reduce the entropy of S (i.e., $H^{\rm S} = \ln 2$) without performing any positive amount of work on S.  We note that, in the conventional thermodynamics, we need a positive amount of work to isothermally decrease the entropy.  In contrast, in the present situation, the mutual information plays the role of the resource to decrease the entropy of S.

We next consider another situation that there is a single logical state $0^{\rm S}$ in S after the feedback, which corresponds to the whole phase-space of S (see also Fig~6 (c)).  In this case, $I = H^{\rm S} = \ln 2$ and $\Delta F_{\rm fb}^{\rm S} =  - \beta^{-1} \ln 2$ hold.  Therefore, inequality~(\ref{feedback_cost1}) reduces to $W_{\rm fb} \geq - \beta^{-1} \ln 2$.  In the quasi-static limit, $W_{\rm fb} = - \beta^{-1} \ln 2$ is achieved, where we extract $\beta^{-1} \ln 2$ of work by the feedback control; the mutual information is the resource of the work extraction.  This situation is equivalent to the case of  the conventional Szilard engine~\cite{Szilard}.

\subsection{On Maxwell's demon}

We consider Maxwell's demon as a special example of our general argument.
A typical situation of Maxwell's demon consists of  measurement and feedback control.
In the setup of Sec.~6.2, memory D plays the role of the demon and memory S is the system to be measured and  controlled.
By summing up inequalities~(\ref{meas_cost1}) and (\ref{feedback_cost1}), we obtain
\begin{equation}
W_{\rm meas} + W_{\rm fb} \geq H^{\rm S} - H^{\rm D} + \Delta F_{\rm meas}^{\rm S} + \Delta F_{\rm fb}^{\rm D},
\label{total_work1}
\end{equation}
where the contribution of the mutual information vanishes in the right-hand side.
If we perform the information erasure from D after the feedback control, the total work is given by
\begin{equation}
W_{\rm meas} + W_{\rm fb} + W_{\rm eras} \geq  H^{\rm S} + \Delta F_{\rm meas}^{\rm S},
\label{total_work2}
\end{equation}
where we used the generalized Landauer principle~(\ref{g_Landauer_work}).
Here, $H^{\rm S} + \Delta F_{\rm meas}^{\rm S}$ on the right-hand side can be regarded as an effective free energy of S, which vanishes in, for example, the case of the Szilard engine discussed in Sec.~6.2.  If $H^{\rm S} + \Delta F_{\rm meas}^{\rm S} = 0$ holds, inequality~(\ref{total_work2}) reduces to
\begin{equation}
W_{\rm meas} + W_{\rm fb} + W_{\rm eras}  \geq 0,
\end{equation}
which implies that we cannot extract any work from the entire process.

We stress that, before the information erasure,  the total entropy productions~(\ref{tot_meas}) and (\ref{tot_feedback}) are always nonnegative (i.e., $\Delta S_{\rm tot}\geq 0$) for the individual processes of measurement and feedback. 
This confirms that  measurement and feedback control are individually consistent with the second law of thermodynamics, without considering the information erasure.

As shown in Sec.~6.2,  both of measurement and feedback are logically reversible in typical situations, and can be thermodynamically reversible  in the quasi-static limit.
On the other hand, the information erasure is logically irreversible, while it can be thermodynamically reversible in the quasi-static limit.
Therefore, the entire process of Maxwell's demon including the erasure is logically irreversible, but can be thermodynamically reversible in the quasi-static limit, where $\Delta S_{\rm tot} = 0$ holds for the individual processes of the measurement, feedback control, and information erasure.

\section{Conclusions}

In this paper, we have discussed the relationship between computation and the second law of thermodynamics.  In particular, we have clarified the fundamental relationship between the thermodynamic and logical reversibilities.

In Sec.~2, we have discussed the concept of thermodynamic reversibility.  A physical process is reversible if and only if  its time-reversal is not prohibited by the second law of thermodynamics.  In more precise, a physical process is  thermodynamically reversible if the total entropy production in the whole universe (\ref{total_entropy_production}) is zero during the process (i.e., $\Delta S_{\rm tot} = 0$).  The total entropy production  consists of the increases in the entropy of the system and that of the heat bath.

In Sec.~3, we have discussed the concept of logical reversibility.  A computational process is logically reversible if and only if it is an injection.  The Shannon entropy of the logical states decreases if a computation is irreversible, while it does not change if a computation is reversible.  A typical example of irreversible computation is the information erasure.

In Sec.~4, we have discussed the conventional setup of the information erasure with a binary symmetric memory (Fig.~1).  On the basis of the second law of thermodynamics, we have confirmed the conventional Landauer principle, which states that at least $\beta^{-1} \ln 2$ of heat should be emitted into the heat bath during the information erasure of one bit of information.  We have clarified that the information erasure can be thermodynamically reversible in the quasi-static limit, where the heat emission equals $\beta^{-1} \ln 2$.    The crucial observation here is that the thermodynamic reversibility is related to the entropy production in the whole universe, while the logical reversibility is related only to the change in the entropy of the logical states.  Such a thermodynamically reversible erasure is illustrated in Fig. 2. 

In Sec.~5, we have discussed the general theory of thermodynamics of computation.  In particular, we have derived a generalized Landauer principle (\ref{g_Landauer2}).  In the case of an asymmetric memory such as Fig.~3 (b) and Fig.~4 (a), the lower bound of the heat emission should be modified from the original Landauer bound;  the second term on the right-hand side of (\ref{g_Landauer2}) originates from the asymmetry of the memory.  In fact, the amount of the heat emission is determined by the increase in the entropy of the heat bath, while the change in the entropy of the logical states of the memory can be compensated for not only by the change in the entropy of the heat bath, but also by that of the internal physical states of the memory if it is asymmetric.  We have clarified the general relationship between the thermodynamic and logical reversibilities, which is summarized in Sec.~5.4; any logically irreversible computation can be thermodynamically reversible in the quasi-static limit.

In Sec.~6, we have discussed the case that there are two memories, and derived a generalized Landauer principle for two memories, which includes the mutual information.  In particular, we have discussed measurement and feedback control, which play the crucial roles in the case of Maxwell's demon.

Our results quantitatively extend the conventional Landauer principle to general situations including asymmetric memories, and clarify the subtle and fundamental relationship between the thermodynamic and logical reversibilities.  The derived inequalities are based on the second law of thermodynamics, where the information contents are included as well as thermodynamic quantities.  Thermodynamics of computation would serve as the foundation to analyze computation at the level of thermal fluctuations with, for example, artificial nanodevices and biological nanomachines.

\ack

The author is grateful to Masahito Ueda, Jordan M. Horowitz, and Juan M. R. Parrondo for valuable discussions, and to Naoto Shiraishi for carefully reading the manuscript.
This work was supported by JSPS KAKENHI Grant Nos. 25800217 and  22340114,  and  by Platform for Dynamic Approaches to Living System from MEXT, Japan.

\end{document}